\documentclass[fleqn,usenatbib]{mnras}

\usepackage[T1]{fontenc}

\DeclareRobustCommand{\VAN}[3]{#2}
\let\VANthebibliography\thebibliography
\def\thebibliography{\DeclareRobustCommand{\VAN}[3]{##3}\VANthebibliography}

\usepackage{graphicx}	% Including figure files
\usepackage{amsmath}	% Advanced maths commands
\usepackage{amssymb}	% Extra maths symbols
\usepackage{float}
\usepackage{newtxtext,newtxmath}
\title[Non-equilibrium: CO and CII emission]{CO and [CII] line emission of molecular clouds -- the impact of stellar feedback and non-equilibrium chemistry\\
}
\author[S. Ebagezio et al.]{
S. Ebagezio,$^{1}$\thanks{E-mail: ebagezio@ph1.uni-koeln.de}
D. Seifried,$^{1,2}$
S. Walch,$^{1,2}$
P.~C. N\"urnberger,$^{1}$
T.-E. Rathjen,$^{1}$
T.~Naab$^{3}$
\\
$^{1}$Universit\"at zu K\"oln, I. Physikalisches Institut, Z\"ulpicher Str.~77, 50937 K\"oln, Germany\\
$^{2}$Center for Data and Simulation Science, University of Cologne, Germany, https://cds.uni-koeln.de\\
$^{3}$Max Planck Institute for Astrophysics, Karl-Schwarzschild-Str.~1, 85748 Garching, Germany
}

\date{Accepted XXX. Received YYY; in original form ZZZ}

\pubyear{2022}

\begin{document}
\label{firstpage}
\pagerange{\pageref{firstpage}--\pageref{lastpage}}
\maketitle

\begin{abstract}
We analyse synthetic $^{12}$CO, $^{13}$CO, and [CII] emission maps of simulated molecular clouds of the SILCC-Zoom project.
We use simulations of hydrodynamical and magnetohydrodynamical clouds, both with and without stellar feedback, including an on-the-fly evolution of seven chemical species including H$_2$, CO, and C$^+$. We introduce a novel post-processing of the C$^+$ abundance using \textsc{Cloudy}, necessary in HII regions to account for higher ionization states of carbon due to stellar radiation. With this post-processing routine, we report self-consistent synthetic emission maps of [CII] in and around feedback bubbles. Within the bubbles a consistent fraction of C$^+$ is actually further ionized into C$^{2+}$ and therefore they are largely devoid of emission [CII], as recently found in observations. The C$^+$ mass is only poorly affected by stellar feedback but the [CII] luminosity increases by $50 - 85$ per cent compared to reference runs without feedback due to the increase of the excitation temperature. Furthermore, we show that, for both $^{12}$CO and $^{13}$CO, the luminosity ratio, $L_\rmn{CO}/L_\rmn{[CII]}$, averaged over the entire cloud, does not show a clear trend and can therefore \textit{not} be used as a reliable measure of the H$_2$ mass fraction or the evolutionary stage of clouds. We note a monotonic relation between the $I_\rmn{CO}/I_\rmn{[CII]}$ intensity ratio and the H$_2$ mass fraction for individual pixels of our synthetic maps, but with a too large scatter to reliably infer the mass fraction of H$_2$. Finally, we show that assuming chemical equilibrium results in an overestimation of H$_2$ and CO masses by up to 110 and 30 per cent, respectively, and in an underestimation of H and C$^+$ masses by 65 and 7 per cent, respectively.  In consequence, $L_\rmn{CO}$ would be overestimated by up to 50 per cent, and $L_\rmn{C[II]}$ be underestimated by up to 35 per cent. Hence, the assumption of chemical equilibrium in molecular cloud simulations introduces intrinsic errors of a factor of up to $\sim2$ in chemical abundances, luminosities and luminosity ratios.
\end{abstract}


%%%%%%%%%%%%%%%%%%%%%%%%%%%%%%%%%%%%%%%%%%%%%%%%%%

%%%%%%%%%%%%%%%%% BODY OF PAPER %%%%%%%%%%%%%%%%%%

\section{Introduction}\label{sec:introduction}

Molecular clouds (MCs) are defined as those regions of the interstellar medium (ISM) where hydrogen exists predominantly in its molecular form, H$_2$. Due to the absence of dipole moment and low temperatures, typically of a few 10 K, H$_2$ is not directly observable in MCs. Nevertheless, information on the abundance and distribution of H$_2$ are of great importance because they allow us to identify the star formation sites in MCs. H$_2$ is observed only indirectly by means of dust continuum \citep[see e.g.][]{Bot2007} and molecules which trace its presence. The most used molecule to trace H$_2$ in MCs is CO \citep[e.g.][]{Wilson1970, Scoville1975, Larson1981, Solomon1987, Dame2001, Bolatto2013, Dobbs2014}. In order to infer the amount of H$_2$ from the observations of CO, a conversion factor $X_\rmn{CO}$ from the observed CO luminosity into a H$_2$ column density has been established \citep[see e.g.][]{Scoville1987, Dame1993, Strong1996, Melchior2000, Lombardi2006, Nieten2006, Smith2012, Ripple2013, Bolatto2013}. The standard value of $X_\rmn{CO}$ in the Milky Way is commonly assumed to be $X_\rmn{CO} = 2 \times 10^{20}$ cm$^{-2}$ K$^{-1}$ km$^{-1}$ s, but there is plenty of evidence that the actual value strongly depends on the environmental conditions \citep{Glover2011, Shetty2011a, Bolatto2013, Gong2020, Seifried2020} or the cloud's evolutionary stage \citep{Borchert2022}. Furthermore, because of this strong dependence, it is used to assess the total amount of H$_2$ in a cloud, but it cannot be easily used on sub-pc scales \citep[e.g.][]{Bisbas2021}. More in general, CO is not a perfect tracer for H$_2$ because of (i) the presence of CO-dark areas, which may contain a significant amount of H$_2$, but almost no CO \citep[see for instance][]{Lada1988, vanDishoeck1988, Grenier2005, Glover2011, Glover2016, Seifried2020}, and (ii) the optical thickness of CO in denser regions, which break the quantitative relation between the CO luminosity and the H$_2$ mass \citep[e.g.][]{Seifried2020, Bisbas2021}. 

Other chemical species are also used to assess the H$_2$ abundance in the clouds: neutral carbon emission has been studied in this context and an $X_\rmn{C}$-factor \citep{Papadoupolos2004, Offner2014}, defined in an analogous way than $X_\rmn{CO}$, has been used to assess the abundance of H$_2$ in MCs. The value of $X_\rmn{C}$, however, also depends on the clouds' environment \citep{Offner2014, Bisbas2021}. C$^+$ is another frequently abundant form of carbon, which has been studied intensively in MCs and is one of the main coolants of the ISM \citep[][and many more]{Tielens1985, Stacey1991, Stutzki2001, Rollig2006, Ossenkopf2013, Appleton2013, Lesaffre2013, Beuther2014, Pineda2013, Pineda2014, Klessen2016}. It is most abundant in photo-dissociation regions \citep{Ossenkopf2013} and in shock fronts \citep{Appleton2013, Lesaffre2013}. Some studies \citep[e.g.][]{Velusamy2014, Franeck2018} suggest that C$^+$ is a tracer of some CO-dark areas of the clouds. However, a  reliable relation between the C$^+$ emission and the H$_2$ abundance is difficult to establish, because most of the [CII] emission comes from regions which are predominantly atomic \citep{Franeck2018}. 

 The formation and evolution of MCs has been studied with numerical simulations in a large number of recent works \citep[e.g.][and many more]{Dobbs2013, Smith2014, Gatto2015, Li2015,  Walch2015, Ibanez2016, Padoan2016, Seifried2017, Kim2018}. Chemistry treatment is generally performed in two possible ways: one option is to first run the simulations without considering the chemical composition of the clouds and then post-process the chemistry assuming equilibrium \citep[e.g.][]{Gong2018, Gong2020, Li2018, Keating2020}. Post-processing the chemistry enables the use of complex networks, but the assumption of chemical equilibrium is necessary and,as a consequence, the H abundance is usually underestimated while H$_2$ is overestimated \citep{Chiayu2021,Borchert2022,Seifried2022}. Conversely, some other simulations include a treatment of molecule formation with a non-equilibrium chemical network \citep{Clark2012, Smith2014a, Smith2014b,  Seifried2016, Walch2015, Hu2016, Hu2017, Smith2020, Chiayu2021, Valdivia2016, Lahen2020}. This usually implies the usage of simpler networks but the assumption of chemical equilibrium is avoided. Recently, non-equilibrium chemistry has been joined with high-resolution simulations. For instance, in the SILCC-Zoom project \citep{Seifried2017, Seifried2020, Haid2019}, the formation of molecular clouds is followed from spatial scales of several hundred parsec down to $\sim 0.1$~pc. These simulations serve as a basis of this publication.

In this paper, we produce synthetic observations of these simulated MCs using the \textsc{RADMC-3D} radiative transfer code \citep{Dullemond2012} in order to investigate (i) the usability of the CO/[CII] emission line ratio as an alternative tracer (to X$_\rmn{CO}$) for the prevailing H$_2$ gas mass, and (ii) its potential of being used as an indicator of MC evolution. We also shed light on the role of the assumption of equilibrium chemistry on the emission of CO and [CII]. 

This paper is structured as follows: in Section \ref{sec:numerics} we describe the numerical methods, which we use to run the simulations and the radiative transfer calculations. In Section \ref{sec:results} we describe the overall aspect of the simulations, the corresponding synthetic observations, the $X_\rmn{CO}$ factor, and the line ratios, considering both the total luminosity and the intensity from single pixels. Then, we discuss our results and we analyse the importance of the equilibrium chemistry in Section \ref{sec:discussion}. Finally, we summarise our results in Section \ref{sec:conclusions}.

\section{SILCC-Zoom simulations}\label{sec:silcc-zoom}

The simulated MCs we use in this paper are part of the SILCC-Zoom project \citep{Seifried2017}. The zoom-in simulations are performed within the SILCC project \citep[see][for details]{Walch2015, Girichidis2016}.

The SILCC setup models a region of a stratified galactic disc. The rectangular box measures 500~pc $\times$ 500~pc $\times \pm 5$~kpc, and uses periodic boundary conditions in \textit{x}- and \textit{y}-direction, while outflow boundary conditions are applied in the \textit{z}-direction. The simulations are performed with the adaptive mesh refinement (AMR) code \textsc{FLASH} 4.3 \citep{Fryxell2000, Dubey2008} and use, for the hydrodynamics (HD) runs, a solver described in \citet{Bouchut2007, Waagan2009}, which guarantees positive entropy and density. The magnetohydrodynamical (MHD) runs use an entropy-stable solver \citep{Derigs2016, Derigs2018}. We model the chemical evolution of the ISM using a chemical network for H$^+$, H, H$_\rmn{2}$, C$^+$, O, CO, and e$^\rmn{-}$ \citep{Nelson1997, Glover2007a, Glover2007b, Glover2010}, hereafter NL97, which also follows the thermal evolution of the gas including the most important heating and cooling processes. We assume solar metallicity with elemental abundances of carbon and oxygen relative to hydrogen of $1.4 \times 10^{-4}$ and $3.16 \times 10^{-4}$, respectively \citep{Sembach2000}. The ISM is embedded in an interstellar radiation field (ISRF) of $G_0 = 1.7$ in units of \citet{Habing1968}, that is in line with \citet{Draine1978}. The cosmic ray ionization rate (CRIR) is set to $3 \times 10^{-17} \text{ s}^{-1}$ with respect to atomic hydrogen. The gas self-gravity as well as the local shielding of the gas from the ISRF is treated with an Octtree-based algorithm described in \citet{Wunsch2018}.

For the magnetized runs, we initialize a magnetic field $\mathbf{B}$ along the \textit{x}-direction as 
\begin{equation}\label{eq:b_field}
B_\rmn{x} = B_\rmn{x,0} \sqrt{\rho(z)/ \rho_0} \; ,
\end{equation}
where $B_\rmn{x,0} = 3 \, \mu \text{G}$ is in accordance to recent observations \citep[e.g.][]{Beck2013}, $\rho_0 = 9 \times 10^{-24} \, \text{g cm}^{-3}$ \citep[see][for more details]{Walch2015}, and $\rho(z)$ is the initial Gaussian density distribution, with $z$ being the distance from the galactic midplane.

\begin{table}
	\caption{Overview of the simulations giving the run name, the zoom time $t_0$, the run type (hydrodynamical, HD, or magnetohydrodynamical, MHD), and stellar feedback}\label{tab:t_zoom}
	\begin{tabular}{|c|c|c|c|}
		\hline
		run name & $t_0$ [Myr] & run type  & feedback \\
		\hline
	    MC1-HD-noFB & 11.9 & HD & no \\
	    MC1-HD-FB & 11.9 & HD & yes \\
	    MC2-HD-noFB & 11.9 & HD & no \\
	    MC2-HD-FB & 11.9 & HD & yes \\
	    MC1-MHD-noFB & 16.0 & MHD & no \\
	    MC1-MHD-FB & 16.0 & MHD & yes \\
	    MC2-MHD-noFB & 16.0 & MHD & no \\
	    MC2-MHD-FB & 16.0 & MHD & yes \\
		\hline
	\end{tabular}
\end{table}

Up to a time $t_0$ after the beginning of the simulations, supernova explosions drive turbulence. The rate at which the supernovae are injected is based on the Kennicutt -- Schmidt relation, relating the disc's surface density (here 10 M$_{\sun}$ pc$^{-2}$) with a typical star formation rate surface density. The latter is translated into a supernovae rate by assuming a standard initial mass function. We refer to \citet{Walch2015} and \citet{Girichidis2016} and references therein for details. At $t_0$ the further injection of background supernovae is stopped and local gas overdensities, i.e., the regions where MCs are about to form, are already visible. We select a few of these ``zoom-in'' regions and continue the simulations allowing for a resolution up to 0.12 pc in those regions. The typical side length of the zoom-in regions is about 100 pc.  We consider two purely hydrodynamical clouds, which we refer to as MC1-HD and MC2-HD, and two magnetohydrodynamical clouds, MC1-MHD and MC2-MHD. The HD runs are described in detail in \citet{Seifried2017} and the MHD runs in \citet{Seifried2019}. We emphasise that HD and MHD runs refer to different clouds, they are not just the same clouds with/without external magnetic field included. In MHD runs $t_0$ is larger than in the HD runs due to the slower dynamical evolution of the MHD clouds \citep{Walch2015, Girichidis2016}.

We set $t_0 = 11.9$~Myr for the HD runs and $t_0 = $ 16.0~Myr for the MHD runs. We also run these 4 clouds with stellar feedback included. We distinguish throughout the text by indicating, for instance, MC1-HD-noFB and MC1-HD-FB for non-feedback and feedback runs, respectively. An overview of the simulations features is given in Table \ref{tab:t_zoom}. In the feedback runs \citep[see][for a more detailed description]{Haid2019}, we use sink particles to model the formation and the evolution of stars and, as a consequence, of the ionizing radiation feedback by massive stars. The sinks form when the gas density exceeds $1.1 \times 10^{-20}$ g cm$^{-3}$. Further, the nearby gas has to be in a converging flow, gravitationally bound, Jeans unstable, and in a local gravitational potential minimum. Details on this are provided in \citet{Federrath2010}. We also use the sink accretion criteria described there. Next, per 120 M$_\odot$ of accreted mass, a massive star between 9 and 120 M$_\odot$ is created using a random sampling algorithm \citep{Gatto2017}, which follows the initial mass function of \citet{Salpeter1955}.  The radiative feedback relative to each star is treated with TreeRay \citep{Wunsch2021}. The chemical evolution of all HD clouds, as well as MC1-MHD-noFB and MC2-MHD-noFB, in particular for H$_2$ and CO, is reported in \citet{Seifried2020}. 

\section{Post-processing and radiative transfer}\label{sec:twopostprocesssteps}

Before the actual synthetic emission maps are produced (Section~\ref{sec:radmc}) we prepare the necessary input data in two steps. The first step is applied to the entire simulation domain (Section~\ref{sec:other_post_process}), the second is only applied to regions affected by stellar feedback (Section~\ref{sec:postprocess}). Unless stated otherwise, all the results shown in the remainder of the paper include the post-processing described in the following.

\subsection{Data preparation for the radiative transfer}\label{sec:other_post_process}

In the first step we apply a pipeline developed by P.~C.~N\"urnberger\footnote{https://bitbucket.org/pierrenbg/flash-pp-pipeline/src/master/} to the entire simulation domain. It converts the \textsc{FLASH} simulation data into input files for \textsc{RADMC-3D} \citep[][used for the radiative transfer]{Dullemond2012} and incorporates further physical processes: (i) CO freeze-out, (ii) $\rmn{C}^{\rmn{+}} \rightarrow \rmn{C}^{\rmn{2+}}$ thermal ionization, (iii) splitting of H$_2$ into para- and orho-H$_2$, and (iv) the generation of a microturbulence file. 

\begin{itemize}
    \item [(i)] The CO density after the freeze-out post-processing, $n_\rmn{CO, f}$, is obtained from the original density $n_\rmn{CO}$ following \citet{Glover2016} and references therein: 
    \begin{equation}\label{eq:co_freezeout}
        n_\rmn{CO, f} = n_\rmn{CO} \times \frac{k_\rmn{cr} + k_\rmn{therm}}{k_\rmn{cr} + k_\rmn{therm} + k_\rmn{ads}} \; ,
    \end{equation}
    where 
    \begin{equation}\label{eq:k_cr}
        k_\rmn{cr} = 5.7 \times 10^4 \times \, \text{CRIR} 
    \end{equation}
    is the CR-induced desorption rate of CO from dust grains,
    \begin{equation}\label{k_term}
        k_\rmn{therm} = 1.04 \times 10^{12} \, \text{exp}\left(-\frac{960 \, \rmn{K}}{T_\rmn{d}} \right) \; 
    \end{equation}
    is the thermal desorption rate. Here, $T_\rmn{g}$ is the gas temperature and $T_\rmn{d}$ is the dust temperature. Furthermore, 
    \begin{equation}\label{eq:k_ads}
        k_\rmn{ads} = 3.44 \times 10^{-18} \, \sqrt{T_\rmn{g}} \, (2 n_\rmn{H_2} + n_\rmn{H}) \; 
    \end{equation}
    is the adsorption rate due to collisions between CO and dust grains.
    
    \item[(ii)] The C$^+$ $\rightarrow$ C$^{2+}$ ionization due to collisions induced via thermal motions is implemented following \citet{Sutherland1993}. In cells with a gas temperature $T_\rmn{g} \geq 2 \times 10^4 \text{ K}$ the C$^+$ density is corrected in order to consider such collisions. 
    
    \item[(iii)] We distinguish between the two nuclear spin states of H$_\rmn{2}$, in which the spins of the nuclei are parallel ($\rmn{ortho-H_2}$) or anti-parallel ($\rmn{para-H_2}$). These are given following, \citet{Rachford2009}, by
    \begin{equation}\label{eq:para}
        n\rmn{(para-H_2)} = \frac{n_\rmn{H_2}}{9 \, e^{-170.5 \, \rmn{K} / T_\rmn{g}} + 1} \;
    \end{equation}
    and
    \begin{equation}\label{eq:n_ortho}
        n\rmn{(ortho-H_2)} = n_\rmn{H_2} - n\rmn{(para-H_2)} \; .
    \end{equation}
    If Eq.~\eqref{eq:para} and \eqref{eq:n_ortho} yield to an ortho-to-para ratio larger than 3, we force $n\rmn{(ortho-H_2)}/n\rmn{(para-H_2)} = 3$, with $n\rmn{(ortho-H_2)} + n\rmn{(para-H_2)} = n_\rmn{H_2}$.  
     \item[(iv)] Microturbulence is also included in our radiative transfer calculations. We assume that the microturbulence broadening is as strong as the thermal broadening. Therefore, with $a$ being the line width, we have $a^2 = a^2_\rmn{therm} + a^2_\rmn{turb}$, with 
\begin{equation}\label{eq:microturbulence}
    a_\rmn{therm} = a_\rmn{turb} = \sqrt{\frac{2k_B T_\rmn{g}}{\mu \, m_\rmn{p}}} \; ,
\end{equation}
where $k_B$ is the Boltzmann constant, $\mu = 2.3$ is the assumed mean molecular mass of the gas, and $m_\rmn{p}$ is the proton mass.
    
\end{itemize}

\subsection{Chemical post-processing of stellar feedback regions}\label{sec:postprocess}
This second data preparation step is only applied to regions affected by stellar feedback to properly model the abundance of C$^+$. The NL97 chemical network does not contain any higher ionized states of carbon than C$^{+}$ and the ionization to C$^{2+}$ described before only accounts for thermal ionization. Stellar radiation, however, can cause further ionization of carbon \citep[see e.g.][]{Abel2005}: the stars formed in our simulations have masses greater than, or equal to, 9 M$_\odot$ and are therefore O-type or B-type stars. These stars, whose effective temperature equals or exceeds $24\,000$ K, emit photons with energies larger than the second ionization energy of carbon of 24.4~eV. It is therefore necessary to remove C$^{+}$, which gets further ionized by this radiation, in order to obtain more realistic emission maps of the feedback runs. For this purpose we use a novel approach based on \textsc{Cloudy} \citep{Ferland2017}, which we describe in detail in the following.

\subsubsection{\textsc{Cloudy} database}

We consider 4 parameters provided by the \textsc{FLASH} simulation data as an input for \textsc{Cloudy}. The gas density, $n_\rmn{g}$, the gas temperature, $T_\rmn{g}$, the energy density of ionizing photons, $E_\rmn{ion}$ (converted later to a bolometric luminosity, see below), and the temperature of the star responsible for the ionization, $T_\star$. In order to avoid to run \textsc{Cloudy} for each simulation cell, we create a database beforehand. To do this, we vary the parameters mentioned above over the range of values found in our simulations, summarized in Table~\ref{tab:t_cloudy}. We run one \textsc{Cloudy} model for each possible combination of values, which corresponds to approximately 160,000 models in total.

\textsc{Cloudy} requires for a source of ionizing photons its bolometric luminosity, $L_\rmn{bol}$, and not $E_\rmn{ion}$. Therefore, we convert $E_\rmn{ion}$ into $L_\rmn{bol}$ as follows. Given a defined spectral luminosity $L_\nu$ for a star, $L_\rmn{bol}$ is defined as
\begin{equation}
    L_\rmn{bol} = \int_0^{\infty} L_\nu d \nu \; .
\end{equation}
Similarly, we define the ionizing luminosity $L_\rmn{ion}$ as the spectral luminosity integrated over the frequencies larger than the ionizing frequency of atomic hydrogen, $\nu_\rmn{H} = 13.6\,\rmn{eV}/h$, where $h$ is the Planck constant:
\begin{equation}
    L_\rmn{ion} = \int_{\nu_\rmn{H}}^\infty L_\nu d \nu \; .
    \label{eq:Lion}
\end{equation}
Furthermore, assuming that the emission spectrum of the chosen star (see below how we choose the star) is equal to that of a black body with its temperature $T_\star$, $B_\nu \left(T_\star \right)$, we have 
\begin{equation}\label{eq:L_bol}
    L_\rmn{bol} = L_\rmn{ion} \frac{\int_0^\infty B_\nu \left(T_\star \right) d \nu}{\int_{\nu_\rmn{H}}^\infty B_\nu \left(T_\star \right) d \nu} \; .
\end{equation}
The energy density of ionizing photons, provided in the \textsc{FLASH} simulation data, is related to the ionizing luminosity (Eq.~\ref{eq:Lion}) via
\begin{equation}\label{eq:E_ion}
    E_\rmn{ion} = \frac{1}{c} \, \frac{L_\rmn{ion}}{4 \pi d^2} \; ,
\end{equation}
with $d$ being the distance between the star and the investigated point. Therefore, we can convert $E_\rmn{ion}$ into $L_\rmn{bol}$ via
\begin{equation}
    L_\rmn{bol} = E_\rmn{ion}  \times {4 \pi d^2 c} \times \frac{\int_0^\infty B_\nu \left(T_\star \right) d \nu}{\int_{\nu_\rmn{H}}^\infty B_\nu \left(T_\star \right) d \nu} \, .
    \label{eq:Lbol-cloudy}
\end{equation}

In our \textsc{Cloudy} models we now assume a fixed $d$ of 100 pc. Hence, Eq.~\ref{eq:Lbol-cloudy} gives us the luminosity of a \textit{hypothetical} star at a distance of 100~pc, which would provide the exactly same value for $E_\rmn{ion}$ as the actual star at its real distance due to the attenuation by gas and dust in between. This approach thus limits the parameter range to be covered by the \textsc{Cloudy} database to 4 dimensions without loss of generality. Furthermore, the cells in our AMR simulations have a size of $dx = $ 0.122, 0.244, 0.488, and 0.976~pc. We take $dx$ = 0.976~pc as the depth of the PDR in \textsc{Cloudy}, which automatically includes the results for all smaller cell sizes (see below). In addition, our choice of \mbox{$d =$ 100 pc} assures that $dx << d$, that is, the \textsc{Cloudy} models are essentially a plane-parallel PDR which we can then directly map back to the simulation cell.

\textsc{Cloudy} provides the fractional abundance $f_\rmn{X,cloudy}$ of the chemical species X with respect to the total number of hydrogen nuclei, i.e., $f_\rmn{X,cloudy} = n_\rmn{X,cloudy}/n_\rmn{H,tot}$, as a function of the distance from the edge of the slab. We are interested in the mean value over the cell, to which the slab corresponds. As the length of the (quasi plane-parallel) PDR slab is 0.976~pc, in a next step we average the chemical abundance in the PDR slab from 0 to a depth of $dx$ = 0.122, 0.244, 0.488, and 0.976~pc, i.e.~each \textsc{Cloudy} model provides now four values of $f_\rmn{X,cloudy}$, one for each cell size. Hence, at this point we have a database covering the full range of relevant physical parameters and possible cell lengths.

\begin{table}
	\caption{Parameter range used for \textsc{Cloudy} models. We run one model for each combination of parameters. Steps are equally spaced in log-scale.}\label{tab:t_cloudy}
	\centering
	\begin{tabular}{|c|c|c|c|}
		\hline
		Parameter & min & max & \# steps \\
		\hline
        $n_\rmn{g}$ [g cm$^{-3}$] & 10$^{-26}$ & 10$^{-20}$ & 19 \\
        $T_\rmn{g}$ [K] & 10$^{1.5}$ & 10$^{4.5}$ & 19 \\
        $T_\star$ [K] & 10$^{3.5}$ & 10$^{5.5}$ & 21 \\
        $E_\rmn{ion}$ [erg cm$^{-3}$] & 10$^{-18}$ & 10$^{-8}$ & 21 \\
		\hline
	\end{tabular}
\end{table}

\subsubsection{Calculation of the new C$^+$ abundance}

In a next step we now post-process the \textsc{RADMC-3D} input file concerning C$^+$ on a cell-by-cell basis as follows.
\begin{itemize}
    \item We check whether $E_\rmn{ion} > 0$. If this is not the case, we skip the following points and do not post-process the cell. Otherwise,
    \item In order to estimate which star contributes most to the flux of ionizing photons at the considered cell, we loop over all stars and compute the unattenuated flux $F_i$ from to the $i$-th star reaching the cell:
    \begin{equation}
        F_i = L_{\rmn{bol,}i}/4\pi d_i^2 \, .
    \end{equation}
   Here, $d_i$ is the \textit{actual} distance between the cell and the $i$-th star and $L_{\rmn{bol,}i}$ the \textit{actual} bolometric luminosity of the $i$-th star obtained directly from the simulation data.
    \item We select the $k$-th star, with temperature $T_{\star,k}$, for which \hbox{$F_k = \max(F_i)$}. For consistency, we check that $d_k \leq 2d_\rmn{min}$, with $d_\rmn{min} = \min(d_i)$. This is always the case in the simulations presented in this paper. In the following we assume that star $k$ is solely responsible for all the ionizing photons.
    \item Next, we take the input values $n_\rmn{g}$, $T_\rmn{g}$, $T_{\rmn{\star},k}$, $E_\rmn{ion}$, and the cell size $dx$ and interpolate the database for $f_\rmn{C^+,cloudy}$ created before using \textsc{LinearNDInterpolator}, which is part of \textsc{Python's NumPy} library, to obtain the updated value $f_\rmn{C+}$ for the given cell.
    \item Finally, we replace the original C$^+$ number density of the cell with $f_\rmn{C+} \times n_\rmn{H,tot}$, where $n_\rmn{H,tot}$ is the total hydrogen nuclei number density.

\end{itemize}

\subsection{Radiative transfer}\label{sec:radmc}
We create synthetic emission maps of $^{12}$CO (1 $\rightarrow$ 0) at 2600 $\rmn{\mu}$m, $^{13}$CO (1 $\rightarrow$ 0) at 2720 $\rmn{\mu}$m, and [CII] at 158 $\rmn{\mu}$m for all 8 clouds (see Table \ref{tab:t_zoom}) at three different evolutionary stages separated by 1 Myr for lines of sight (LOS) along the $x$-, $y$-, and $z$-axis. The emission maps have the same resolution as the simulations, i.e.~0.12 pc. The radiative transfer, which is needed to obtain synthetic emission maps of the simulated clouds, is performed using the \textsc{RADMC-3D} software, an open-source, 3D radiative transfer code\footnote{\url{http://www.ita.uni-heidelberg.de/~dullemond/software/radmc-3d/}} \citep{Dullemond2012}. We include microturbulence (Eq.~\ref{eq:microturbulence}) and use the Large Velocity Gradient approximation \citep[LVG;][]{Ossenkopf1997, Shetty2011a, Shetty2011b} for calculating the level population. In order to capture the contribution of Doppler-shifted emission, we consider a velocity range of \hbox{$\pm 20$ km s$^\rmn{-1}$}, centred around the selected rest frequency. We divide this range into 201 equally spaced velocity channels, corresponding to a spectral resolution of \hbox{d$v = 0.2$ km s$^\rmn{-1}$}, which is well suited for the studied molecular clouds \citep{Franeck2018}.

Performing the radiative transfer calculations using the LVG approximation means that we do not assume Local Thermal equilibrium (LTE). Therefore, we must specify explicitly the collisional rates for CO and C$^+$. We take the data from the Leiden Atomic and Molecular DAtabase\footnote{\url{https://home.strw.leidenuniv.nl/~moldata/}} \citep[LAMDA,][]{Schoier2005}. We consider para-H$_\rmn{2}$, ortho-H$_\rmn{2}$, H, and e$^\rmn{-}$, as collisional partners for C$^\rmn{+}$, and para-H$_\rmn{2}$, ortho-H$_\rmn{2}$, H, and He, for CO. As the rates for CO-He and CO-H collisions are not in the LAMDA database, for this we use the rates from \citet{Cecchi2002} and \citet{Walker2015}, respectively. We emphasise that it is essential to include also He and H as collisional partners as it increases the CO luminosity by $\sim20 - 30$ per cent \citep{Borchert2022}.

We also consider the Cosmic Microwave Background (CMB) using an isotropic black body emission at 2.725 K. During the following analysis we subtract this background from the emission maps before any other step. Each emission map is centered on a rest frequency $\nu_0$. The corresponding brightness temperature for the CMB is given by
\begin{equation}\label{eq:T_cmb}
    T_\rmn{B,CMB} = \frac{h \nu_0}{k_\rmn{B}} \, \frac{1}{e^{h \nu_0 / kT} -1} \; ,
\end{equation}
where $h$ is the Planck constant and $T = $ 2.725 K. Considering the CMB background and then subtracting it, has a negligible impact on [CII] emission maps, but it changes the $^{12}$CO and $^{13}$CO intensity in optically thick areas by up to $\sim20$ per cent. 

\section{Results}\label{sec:results}
\subsection{Overview of the simulations}\label{sec:overview}

\begin{figure*}
    \includegraphics[width=0.99\textwidth]{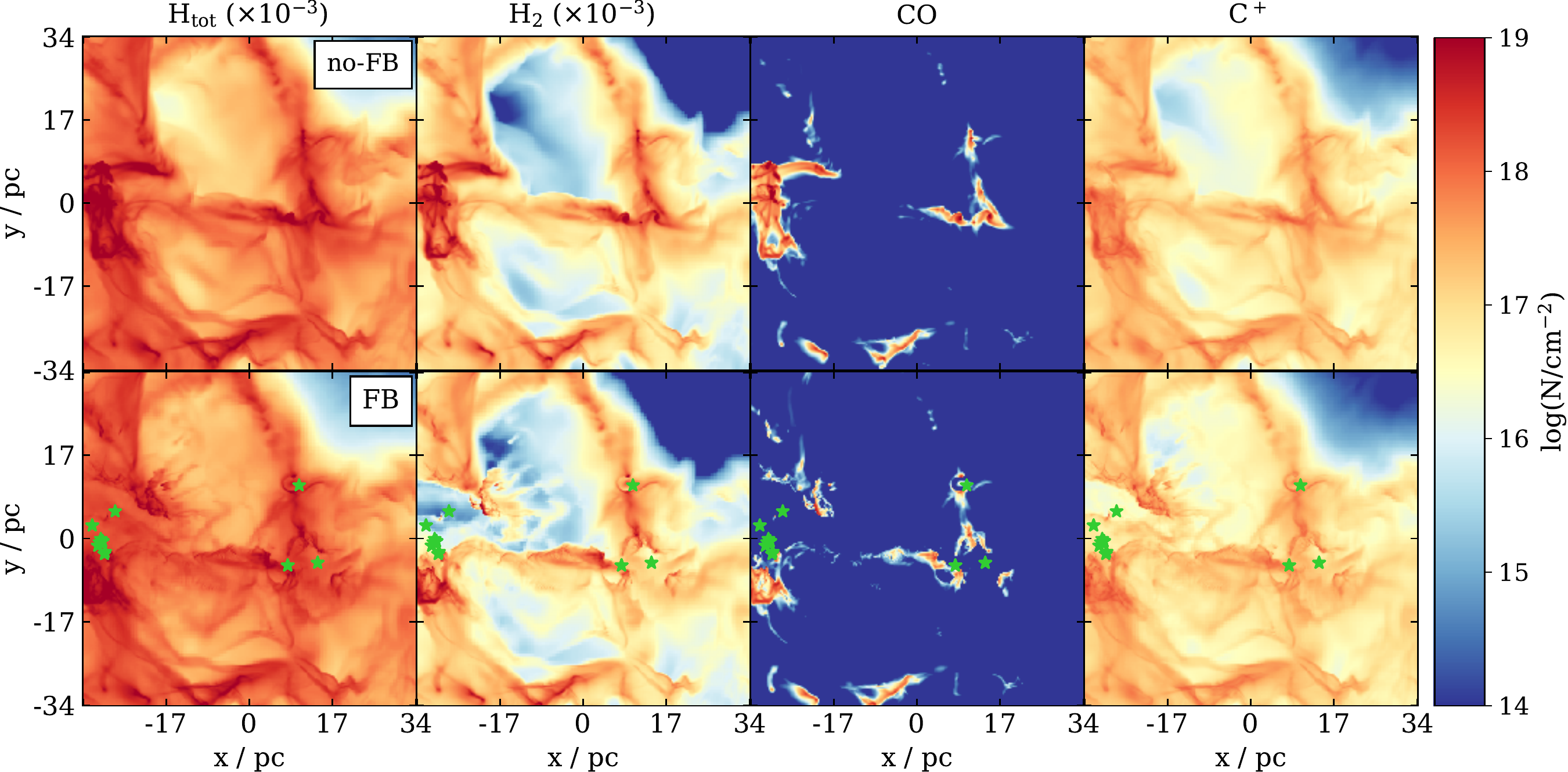}
    \caption{Column density of  H$_\rmn{tot}$, H$_2$, CO, and C$^+$ (from left to right) of the run MC1-HD-noFB (top row) and MC1-HD-FB (bottom row) at $t_\rmn{evol} = 4$ Myr along the $z-$direction. Green symbols mark the positions where stars form. There is an evident nested CO - C$^+$ structure. The H$_2$ distribution is more diffuse than the CO distribution, which leads to a significant CO-dark H$_2$ region. Conversely, the C$^+$ distribution is significantly more diffuse than the H$_2$ distribution. The impact of stellar radiation is evident when comparing particularly dense regions in the noFB run with the corresponding areas in the FB run, as stellar feedback disperses the cloud.}
    \label{fig:col_dens_picture}
\end{figure*}

In the following, we refer to the evolution time $t_\rmn{evol}$ of the clouds as $t_\rmn{evol} = t - t_0$, where $t$ is the time calculated from the very beginning of the SILCC simulation and $t_0$ is the time when the zoom-in simulation starts (see also Table \ref{tab:t_zoom}). As mentioned before, the results shown in the following include the post-processing steps described in Section~\ref{sec:twopostprocesssteps} unless stated otherwise.
	
In Fig.~\ref{fig:col_dens_picture} we show in the top row the H$_\rmn{tot}$, H$_2$, CO, and C$^+$ column densities of MC1-HD-noFB at an evolutionary stage of \hbox{$t_\rmn{evol} = 4$ Myr} along the $z$-axis. In the bottom row we show MC1-HD-FB together with the formed stars. The impact of stellar feedback is evident: it is possible to identify two regions of star formation where stellar radiation disperses the cloud. This is particularly clear when looking at the H$_2$ and CO maps, since the higher temperature and the stellar radiation lead to the dissociation of these molecules. Stellar radiation has also an impact on C$^+$ by further ionizing it to C$^{2+}$ (see Section \ref{sec:postprocess}). Furthermore, comparing the H$_2$ column density with the carbon-bearing species maps, shows that the CO distribution is significantly more compact than the H$_2$ distribution, which leads to the presence of CO-dark H$_2$ regions \citep[see][for details on this]{Seifried2020}. The C$^+$ distribution is significantly more diffuse than the CO distribution, leading to a clearly visible nested CO$-$C$^{+}$ structure.

\begin{figure}
    \centering
    \includegraphics[width=0.65\columnwidth]{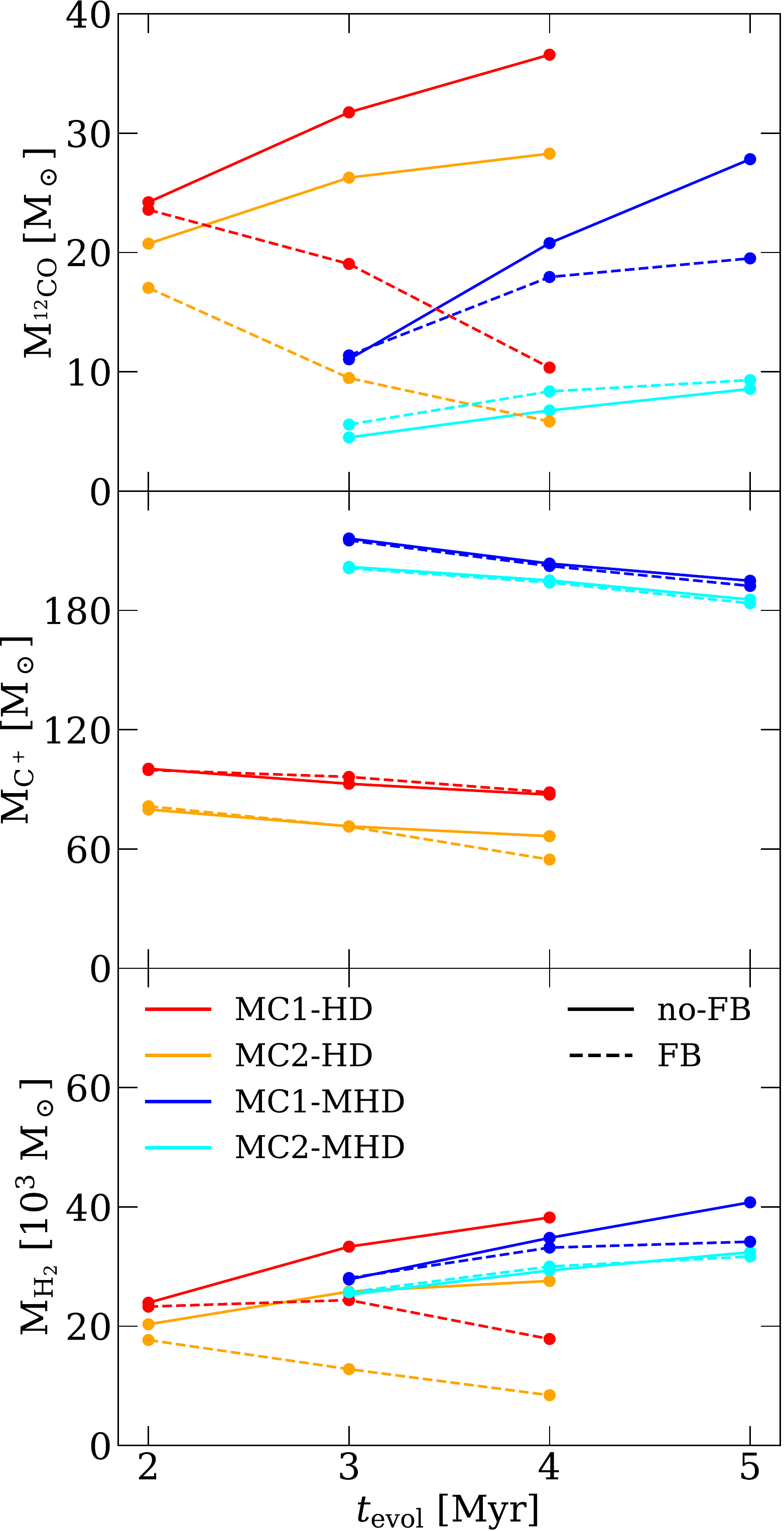}
    \caption{Mass of $^{12}$CO, C$^+$, and H$_2$ (from top to bottom) as a function of time for all four clouds, represented in different colors. The noFB runs are plotted with solid lines, and FB runs in dashed lines. The CO mass 
    rises %raises 
    with time, whereas the C$^+$ mass slowly decreases. HD and MHD clouds show the same trend, but the increase (for noFB runs) of H$_2$ and CO is slowed down in MHD runs with respect to HD runs. Stellar feedback disperses the densest parts of the clouds, where stars form, thus decreasing the mass of CO and H$_2$ over time.}
    \label{fig:abund_over_time}
\end{figure}

Fig.~\ref{fig:abund_over_time} shows the change in CO, C$^+$, and H$_2$ mass as a function of time for the simulated clouds in the zoom-in regions, using selected snapshots separated by 1 Myr in time. In our simulations, $M_\rmn{^{13}CO}$ (not shown here) is fixed to 1/69 of $M_\rmn{^{12}CO}$ \citep{Wilson1999}. For runs without feedback, the CO abundance raises with time, and the C$^+$ abundance slowly decreases. The H$_2$ abundance also rises with time, even though at a lower rate than CO. Both the HD and MHD clouds follow the same trend, but the MHD clouds evolve more slowly: this can be seen, for instance, when considering the C$^+$-to-CO ratio (not shown explicitly in Fig.~\ref{fig:abund_over_time}), which is much higher and more slowly decreasing in the MHD clouds than in the HD clouds. This is due to the inhibiting effect of the magnetic field on formation of dense structure and thus more H$_2$ and CO \citep{Seifried2020}. Stellar feedback reduces the amount of CO and H$_2$ from the onset of star formation. The total amount of C$^+$ is only marginally affected by stellar feedback (due to a partial conversion into C$^{2+}$), although feedback results in a different distribution of C$^+$ (see Fig.~\ref{fig:col_dens_picture}, right column).

\subsection{Synthetic emission maps}\label{sec:synth_obs}
\begin{figure*}
    \centering
    \includegraphics[width=0.99\textwidth]{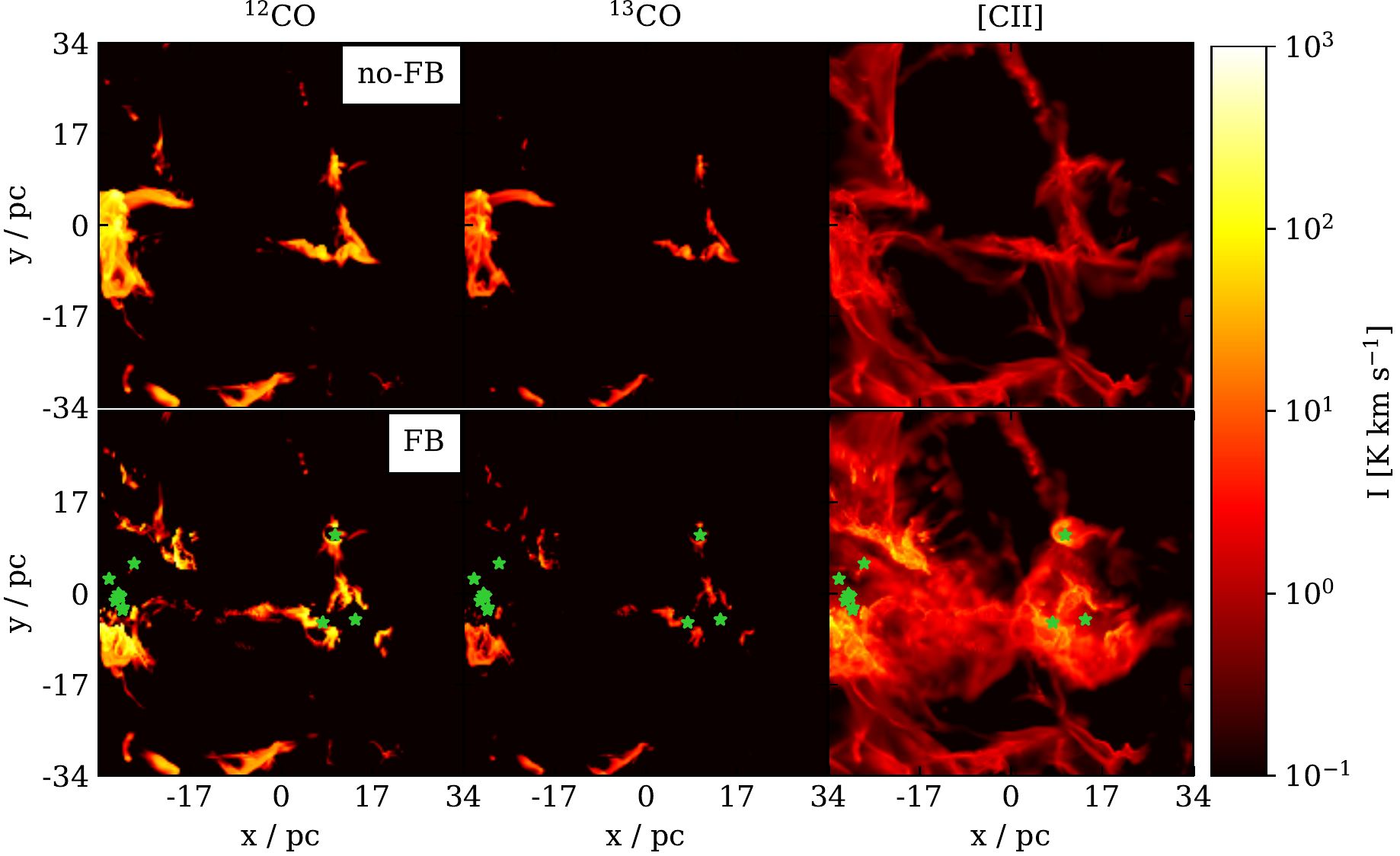}
    \caption{From left to right: integrated emission maps of $^{12}$CO (1 $\rightarrow$ 0), $^{13}$CO (1 $\rightarrow$ 0), and [CII] of MC1-HD without feedback (top row) and with feedback (bottom row) at \hbox{$t_\rmn{evol}$ = 4 Myr.} The CMB background has been subtracted. A nested CO - [CII] is evident in both clouds. HII regions around stars devoid of any or most of the emission are visible in both the [CII] and in CO maps. [CII] intensity is enhanced by an order of magnitude in the rims of the HII regions with respect to the brightest areas in the non-feedback map.}
    \label{fig:emissionmaps}
\end{figure*}

Next, we analyse the emission maps for the same snapshots as in Fig.~\ref{fig:abund_over_time}. In Fig.~\ref{fig:emissionmaps} we show the integrated intensity maps of the $^{12}$CO, $^{13}$CO, and [CII] lines of MC1-HD-noFB (top row) and MC1-HD-FB (bottom row) at $t_\rmn{evol} = 4$ Myr, which corresponds to the column density maps shown in Fig.~\ref{fig:col_dens_picture}. Again, a nested CO - [CII] structure is evident, and stellar feedback removes the CO emission from the expanding bubbles and strongly increases the [CII] emission, in particular from the rims of the bubbles.

Next, we calculate the total luminosity $L$. For this, we first sum the intensity of all pixels to obtain the integrated intensity. Then, $L$ is given by 
\begin{equation}
    L = 4 \pi d^2 \, F \; ,
\end{equation}
where $d$ is the distance of the cloud and $F$ is the total flux derived from the integrated intensity map by adding up the contributions from the total number of pixels, $n$: 
\begin{equation}
    F = \sum_{i = 1}^n I_i \, A_\rmn{pixel} \, .
\end{equation}
Here, $A_\rmn{pixel}$ is the area of the pixels in steradians given by 
\begin{equation}
    A_\rmn{pixel} = \left( \arctan \left(\frac{a}{d} \right) \right)^{2} \; ,
\end{equation}
with $a$ being the side length of the pixel. We note that, due to the small angle approximation, for $a/d \ll 1$ the choice of $d$ is practically irrelevant.

\begin{figure}
    \centering
    \includegraphics[width=0.85\columnwidth]{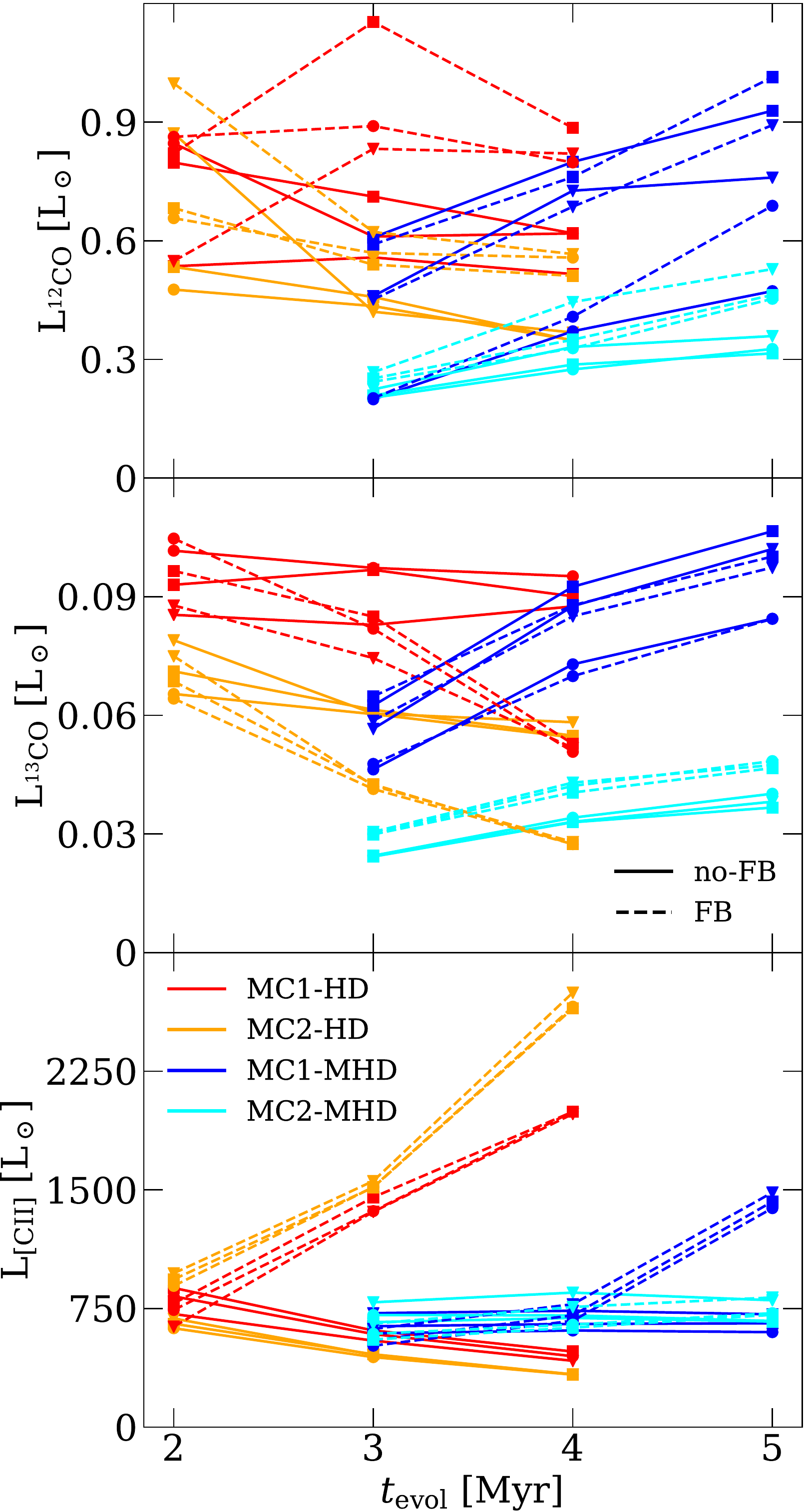}
    \caption{Total luminosity of $^{12}$CO, $^{13}$CO, and [CII] for the investigated clouds. Circles, squares, and triangles indicate values along $z-$, $y-$, and $x-$axis, respectively. The differences in the evolution of $L_\rmn{^{12}CO}$ and $L_\rmn{^{13}CO}$, together with the smaller scatter of $L_\rmn{^{13}CO}$ among the different LOS for the same cloud, indicate that $^{12}$CO is more affected from optical thickness than $^{13}$CO. The significant increase of $L_\rmn{[CII]}$ in the feedback runs is a consequence of the enhanced excitation temperature of C$^+$ due to stellar feedback. This is not the case for MC2-MHD as the stellar feedback plays a minor role for this cloud.} 
    \label{fig:lum_over_time}
\end{figure}

We show the values of $L$ calculated for the three different LOS in Fig.~\ref{fig:lum_over_time}. Optical thickness plays an important role for $L_\rmn{^{12}CO}$, indicated by the fact that the values for the same cloud, but different LOS, differ by up to a factor of a few, whereas in the optically thin case they should be identical. This is also shown by the fact that changes in $M_\rmn{CO}$ (Fig.~\ref{fig:abund_over_time}) are not directly reflected in corresponding changes in $L_\rmn{^{12}CO}$ (e.g. for MC1-HD-noFB). Conversely, the measured $L_\rmn{^{13}CO}$ is less affected by optical thickness: the difference in luminosity for the different LOS is lower than for $^{12}$CO, and $L_\rmn{^{13}CO}$ changes coherently with $M_\rmn{CO}$.

Stellar feedback significantly increases $L_\rmn{[CII]}$ compared to the noFB runs by a factor of 2 -- 7. Only MC2-MHD-FB, which forms the least stellar mass among the clouds we investigate ( 41.9 $M_\odot$ at $t_\rmn{evol}$ = 5 Myr compared to a maximum of 546.2 $M_\odot$ in MC2-HD-FB at 4 Myr), does not show a significant increase in $L_\rmn{[CII]}$. In contrast, $M_\rmn{C^+}$ is practically unchanged between noFB and FB runs (Fig.~\ref{fig:abund_over_time}, middle panel). The increase in $L_\rmn{[CII]}$ is a consequence of the stellar radiation, which heats up the gas and excites the C$^+$ ions: we find that the excitation temperature of [CII] is overall significantly higher in FB (up to 1000 K) than in noFB runs (hardly more than 50 K). In consequence, $L_\rmn{[CII]}$ increases for the FB runs despite a comparable amount of C$^+$ mass between FB and noFB. Most of the [CII] luminosity in these runs comes from the rims of the HII bubbles, as shown for instance in Fig.~\ref{fig:emissionmaps}. \citet{Pineda2013, Pineda2014} claim that 34 -- 70 per cent of the [CII] emission is related to feedback. The factor of 2 -- 7 which we observe for the increase of the [CII] luminosity corresponds to a contribution of 50 -- 85 per cent due to the role of stellar feedback, in rough agreement with the estimate of the aforementioned authors. Moreover, we find that the longer star formation proceeds, the more [CII] increases, thus having an increasingly more important effect on $L_\rmn{[CII]}$. 

\subsection{Feedback-driven [CII] bubbles}\label{sec:bubbles}

\begin{figure*}
    \centering
    \includegraphics[width=0.99\textwidth]{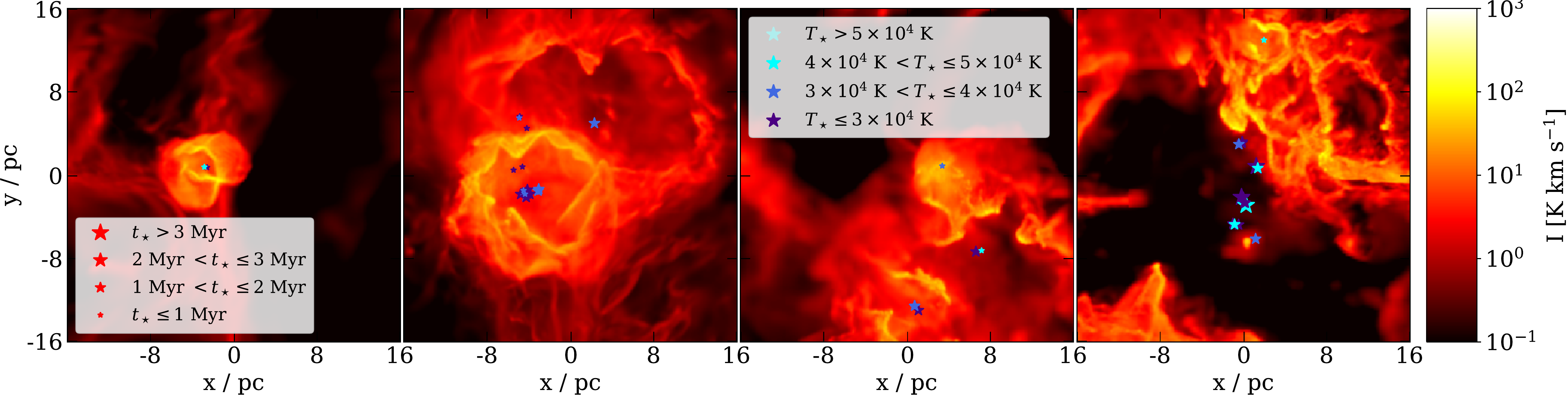}
    \caption{HII regions at different evolutionary stages as seen in [CII]. The individual snapshots are taken from different clouds: from left to right, i) MC2-MHD-FB, $t_\rmn{evol} = 5$ Myr, LOS along the $z-$axis; ii) MC1-MHD-FB, $t_\rmn{evol} = 5$ Myr, LOS along the $y-$axis; iii) MC1-HD-FB, $t_\rmn{evol} = 4$ Myr, LOS along the $y-$axis, and (iv) MC2-HD,  $t_\rmn{evol} = 4$ Myr, LOS along the $x-$axis. The bubbles are ordered, from left to right, from the youngest to the oldest evolutionary stage. Stars formed are superimposed and plotted with different sizes and colors according to their age and temperature. We find than the largest and [CII]-darker bubbles are associated with older stars, whereas smaller and brighter bubbles correspond to younger stars.}
    \label{fig:zoom_on_bubbles}
\end{figure*}

\begin{figure}
    \centering
    \includegraphics[width=0.9\columnwidth]{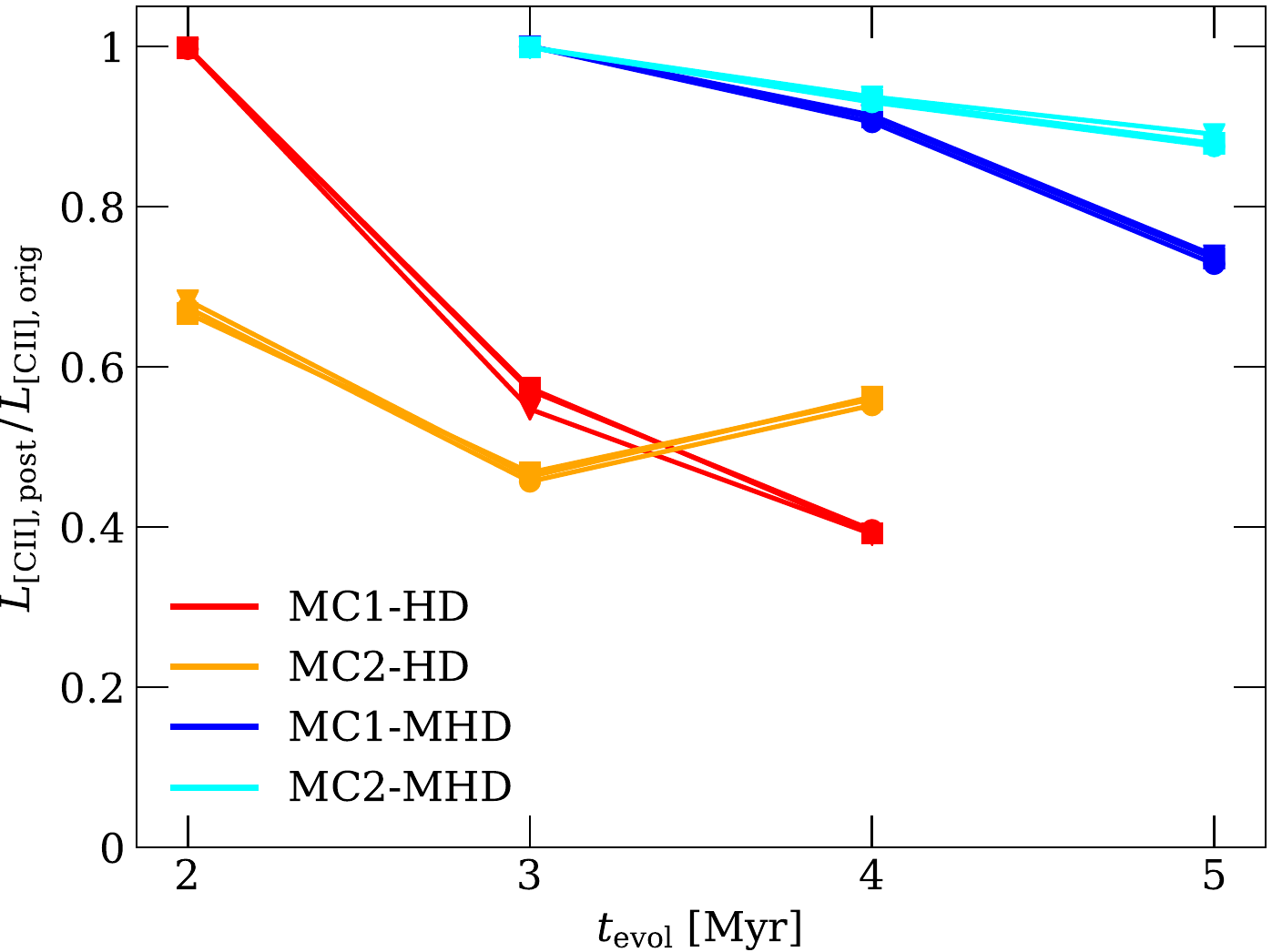}
    \caption{Ratio between the total luminosity of [CII] of the post-processed data, $L_\rmn{[CII], post}$, and the original data, $L_\rmn{[CII], orig}$. Circles, squares, and triangles indicate values along $z-$, $y-$, and $x-$axis, respectively. Post-processing the simulation data removes up to 60 per cent of the [CII] luminosity, especially at later stages where feedback becomes more important. Hence, it is essential to obtain reliable [CII] emission maps.}
    \label{fig:post-process}
\end{figure}

As seen in Fig.~\ref{fig:emissionmaps}, stellar feedback drives some bubbles also visible in [CII], which we investigate in more detail now. For this purpose, in Fig.~\ref{fig:zoom_on_bubbles} we show some examples of HII regions at different evolutionary stages. Stars are superimposed and are characterized with different colors and size according to their age and temperature. In Fig.~\ref{fig:zoom_on_bubbles_comparison} we also show the emission maps of the same regions obtained without operating the post-processing described in Section~\ref{sec:postprocess}. These exhibit a much higher emission coming from the inner regions supporting the importance of the post-processing. We find that some structures become more evident after the post-processing. This is the case of pillars, which can be easily recognised in the maps of MC2-HD-FB. 

In Fig.~\ref{fig:post-process} we summarize the importance of this post-processing step. The total [CII] luminosity decreases by up to 60 per cent if the C$^+$ within the HII regions is post-processed compared to the less realistic non-post-processed case. In general, differences are larger at later evolutionary stages, as the stellar mass and thus stellar radiation intensity increase over time. Hence, the consideration of higher ionization states of carbon is crucial to obtain accurate [CII] intensities stemming from HII regions \citep{Spitzer1978}.

We note that the size of the bubble has a positive correlation, at fixed stellar temperature $T_\star$, with the age $t_\star$ of the stars formed inside. Larger bubbles (corresponding to later evolutionary stages) also show a weaker [CII] emission inside them than smaller bubbles (earlier stages). We also observe a correlation between the star temperature and the lack of [CII] emission inside the bubble. For example, the region around the rather cool star on the top of the map for MC1-HD-FB has still some [CII] emission, whereas the regions in the upper part of the maps of MC2-HD-FB and in the centre of MC2-MHD-FB are almost devoid of any emission. In the latter case $T_\star$ is significantly higher, i.e. the star is able to further ionize C$^+$ as explained in Section~\ref{sec:postprocess}.

At the rims of the bubbles, the [CII] emission is enhanced when compared to even the brightest regions of the noFB runs (Fig.~\ref{fig:emissionmaps}). We emphasise that our findings are in excellent agreement with [CII] bubbles with enhanced emission at the rims and a lack of emission inside found recently in a number of observations \citep[see e.g.][]{Pabst2019, Luisi2021, Tiwari2021}. 

\subsection{The $X_\rmn{CO}$ and $X_\rmn{[CII]}$ factors}\label{sec:x_co}

\begin{figure*}
    \centering
    \includegraphics[width=0.9\textwidth]{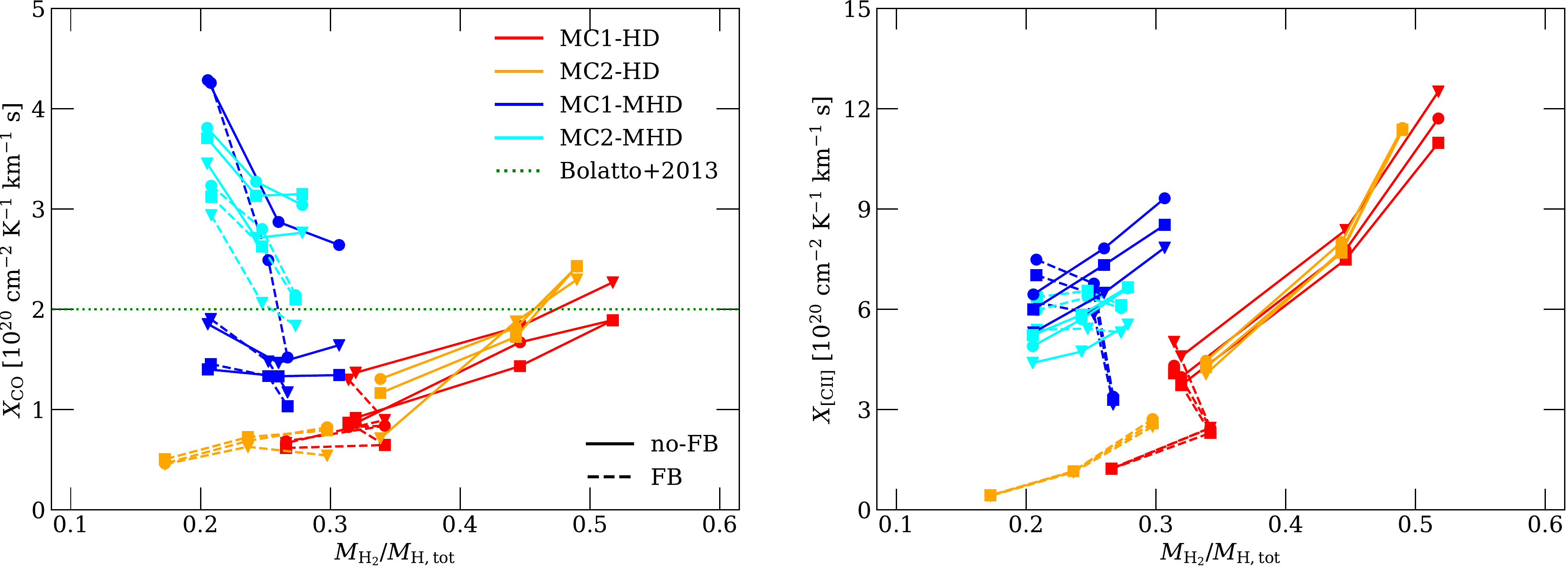}
    \caption{$X_\rmn{CO}$ (left) and $X_\rmn{[CII]}$ (right) of our simulated clouds as a function of the H$_2$ mass fraction. Circles, squares, and triangles indicate values along $z-$, $y-$, and $x-$axis, respectively. Regarding $X_\rmn{CO}$, we find a significant scatter among the different clouds and in time around the reference value for the Milky Way of \hbox{$2 \times 10^{20}$ cm$^{-2}$ K$^{-1}$ km$^{-1}$ s} (dotted line). A similar scatter is found for $X_\rmn{[CII]}$ as well. We see an increase of $X_\rmn{[CII]}$ with $M_\rmn{H_2}/M_\rmn{H, tot}$ for the non-FB runs, wherease the FB runs show no clear trend.}
    \label{fig:x_factor_vs_time}
\end{figure*}

The $X_\rmn{CO}$ factor has been widely studied in literature \citep[see e.g.][]{Scoville1987, Dame1993, Strong1996, Melchior2000, Lombardi2006, Nieten2006, Smith2012, Ripple2013, Bolatto2013}. It is defined as \begin{equation}\label{eq:x_factor}
    X_\rmn{CO} = \frac{N_\rmn{H_2}}{W_\rmn{^{12}CO}} \;,
\end{equation}
where $N_\rmn{H_2}$ is the H$_2$ column density, generally expressed in cm$^{-2}$, and $W_\rmn{^{12}CO}$ the line-integrated intensity, summed over the whole image, expressed in \hbox{K km s$^{-1}$}. Here we calculate $X_\rmn{CO}$ as the average quantity over the entire cloud. It allows to assess the H$_2$ mass of a cloud, given the intensity of the $^{12}$CO ($1 \rightarrow$ 0) transition. The typical $X_\rmn{CO}$ value for the Milky Way is $2 \times 10^{20}$ cm$^{-2}$ K$^{-1}$ km$^{-1}$ s \citep[e.g.][]{Bolatto2013}. We analogously define the $X_\rmn{[CII]}$ factor as 
\begin{equation}\label{eq:x_cii_factor}
    X_\rmn{[CII]} = \frac{N_\rmn{H_2}}{W_\rmn{[CII]}} \;.
\end{equation}
In Fig.~\ref{fig:x_factor_vs_time} we show both factors for our simulations plotted against the mass fraction of H$_2$ \citep[see][for a plot against $t_\rmn{evol}$]{Seifried2020}. We calculate them under the assumption of unresolved clouds, i.e., we first integrate the H$_2$ column density and the CO, or [CII], intensity over the entire area of the zoom-in regions, no matter whether in some pixels the intensity is beyond a minimum observable threshold, and then take the ratio of both values.  

There is a significant scatter of $X_\rmn{CO}$ (left panel of Fig.~\ref{fig:x_factor_vs_time}) around the reference value for the Milky Way. In our clouds it spans from $\sim 0.5$ to \hbox{$\sim 4.5 \times 10^{20}$ cm$^{-2}$ K$^{-1}$ km$^{-1}$ s.} This scatter occurs among different clouds, but also to a smaller extent for the same clouds among different LOS and different $t_\rmn{evol}$. It can in parts be attributed to the fact that a larger CO optical thickness leads to a higher $X_\rmn{CO}$ factor, as the CO intensity does not increase coherently with the H$_2$ mass. Furthermore, differences occur between feedback and non-feedback runs, in particular for HD clouds. Stellar feedback lowers the $X_\rmn{CO}$ factor, as it both slightly enhances the CO emissivity and reduces the H$_2$ mass (see Figs.~\ref{fig:abund_over_time} and \ref{fig:lum_over_time}). 

Moreover, we cannot identify a clear correlation between $X_\rmn{CO}$ and the time evolution, respectively the H$_2$ mass fraction of the clouds. We attribute this to the presence of ``CO-dark'' regions, i.e., molecular gas regions with low or no CO. The amount of CO-dark gas is highly variable in different clouds. As discussed in more detail in \citet{Seifried2020}, the CO-dark gas fractions in our MCs range from 40 to 95 per cent. Indeed, we find that the higher the CO-dark gas fraction, the higher is $X_\rmn{CO}$, e.g. for MC1-HD-noFB and MC2-HD-noFB the CO-dark gas fraction is $\sim 40$ per cent and \hbox{$X_\rmn{CO} = 1 - 2 \times 10^{20}$ cm$^{-2}$ K$^{-1}$ km$^{-1}$ s}, whereas for MC1-MHD-noFB and MC2-MHD-noFB the CO-dark gas fraction is 60 -- 95 per cent, and the associated X$_\rmn{CO}$ is \hbox{$1.5 - 4 \times 10^{20}$ cm$^{-2}$ K$^{-1}$ km$^{-1}$ s}. We note that the X$_\rmn{CO}$ values calculated in \citet{Seifried2020} are slightly different, as there they were calculated considering only the pixel with a CO intensity above a minimum threshold of \hbox{0.1 K km s$^{-1}$.} 

The $X_\rmn{[CII]}$ factor (right panel of Fig.~\ref{fig:x_factor_vs_time}) exhibits a lower scatter between the different LOS of the same cloud, which we attribute to the somewhat lower optical depths in the case of [CII]. Nevertheless, the scatter among different clouds is again significant, with $X_\rmn{[CII]}$ values ranging from 0.5 to 12 $\times 10^{20}$ cm$^{-2}$ K$^{-1}$ km$^{-1}$ s. This means that such $X_\rmn{[CII]}$ is far from being constant and therefore it is difficult to use it as a reliable conversion factor to obtain the H$_2$ mass. Also \citet{Franeck2018} claim that [CII] is not a good tracer of molecular hydrogen, as most of the [CII] intensity comes indeed from atomic hydrogen gas, and not from molecular part. Hence, the monotonic increase in $X_\rmn{[CII]}$ for the noFB runs is mainly due to the increase in H$_2$ mass, while the [CII] luminosity stemming from the outskirts of the clouds remains largely constant (see Figs.~\ref{fig:abund_over_time} and~\ref{fig:lum_over_time}).
The FB runs, conversely, do not exhibit a clear relation because the stellar feedback both inhibits the formation of H$_2$ (Fig.~\ref{fig:abund_over_time}) and enhances the [CII] intensity (Fig.~\ref{fig:lum_over_time}).

\subsection{The global CO/[CII] line ratio}\label{sec:line_ratio}

\begin{figure*}
    \centering
    \includegraphics[width=0.99\textwidth]{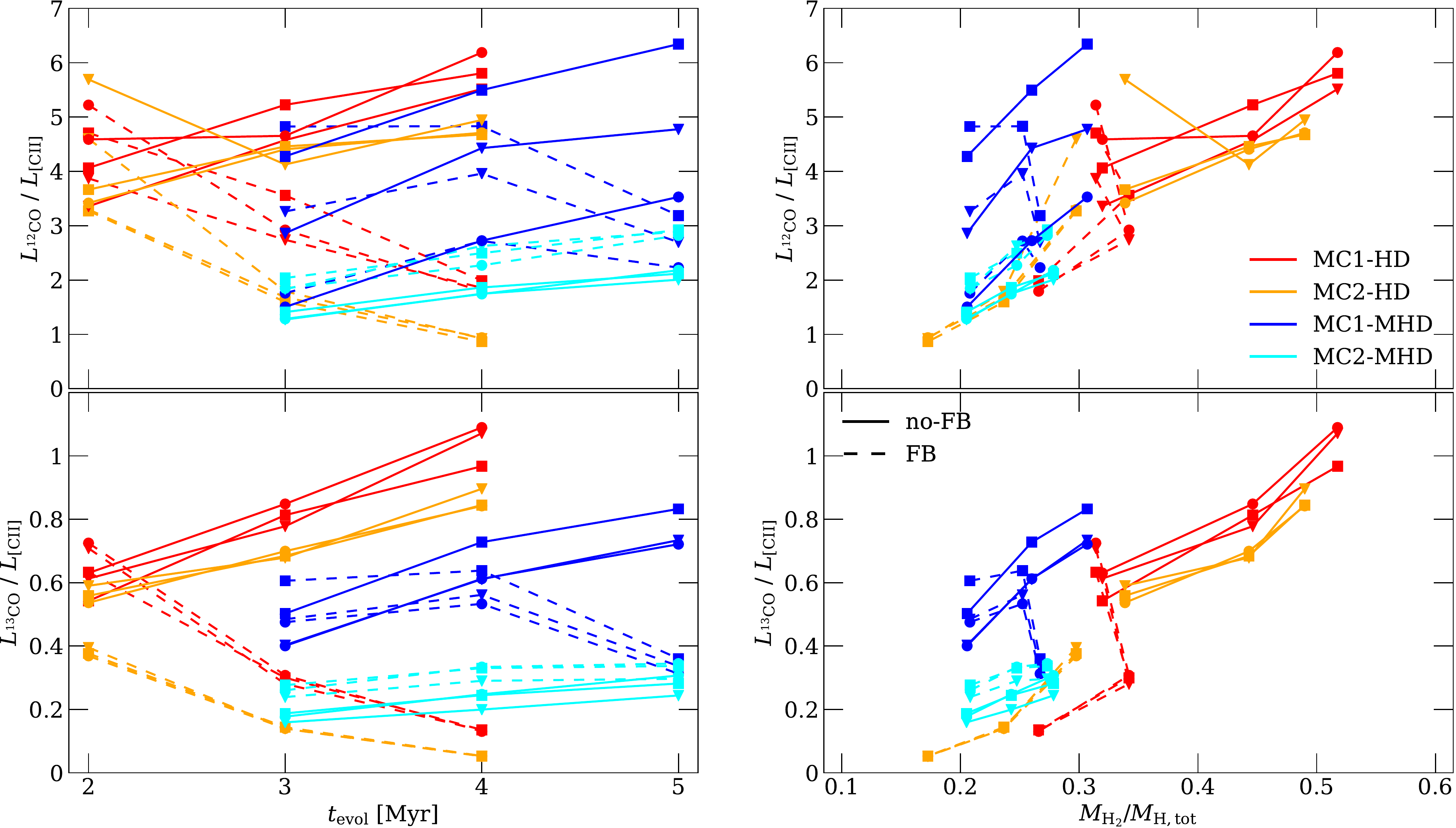}
    \caption{$L_\rmn{^{12}CO}/L_\rmn{[CII]}$ (top row) and $L_\rmn{^{13}CO}/L_\rmn{[CII]}$ (bottom row) as a function of $t_\rmn{evol}$ (left column) and $M_\rmn{H_2}/M_\rmn{H_{tot}}$ (right column). For all noFB runs the line ratio increases with $t_\rmn{evol}$. For all FB runs it decreases, with the only exception of MC2-MHD-FB, which has less dense gas. 
    Altogether, there is no clear trend of the luminosity ratio with either $t_\rmn{evol}$ and $M_\rmn{H_2}/M_\rmn{H,tot}$. In addition, there is a large scatter for a given evolutionary stage. The scatter among different LOS is reduced when considering $^{13}$CO, due to its smaller optical thickness.} 
    \label{fig:line_ratio_global_fit}
\end{figure*}

The large scatter in $X_\rmn{CO}$ makes it difficult to use it to reliably estimate the H$_2$ mass for individual clouds \citep{Bolatto2013, Seifried2020, Madden2020, Hu2022}. We therefore consider another possible estimator: We take the total luminosities integrated over the entire cloud, $L_\rmn{^{12}CO}$, $L_\rmn{^{13}CO}$, and $L_\rmn{[CII]}$, and investigate the relation between $L_\rmn{^{12}CO}$/$L_\rmn{[CII]}$, $L_\rmn{^{13}CO}$/$L_\rmn{[CII]}$, and the H$_2$ mass fraction of the clouds, $M_\rmn{H_2} / M_\rmn{H,tot}$. 

Fig.~\ref{fig:line_ratio_global_fit} shows $L_\rmn{^{12}CO}$/$L_\rmn{[CII]}$ and $L_\rmn{^{13}CO}$/$L_\rmn{[CII]}$ as a function of time (top row) and as a function of the H$_2$ mass fraction $M_\rmn{H_2}/M_\rmn{H,tot}$ (bottom row). In the Appendix (Figs.~\ref{fig:line_ratio_vs_mass} and \ref{fig:line_ratio_erg}) we also show the line ratios as a function of the H$_2$ mass and the [CII]/CO ratio calculated with luminosities given in units of erg s$^{-1}$ cm$^{-2}$ in order to allow for an easier comparison with observational literature. We note that we relate the luminosity ratio to $M_\rmn{H_2}/M_\rmn{H,tot}$ rather than $M_\rmn{H_2}$, as the latter depends on the size of the cloud, i.e.~it is an extensive quantity and not an intensive one as the mass fraction. 

We find that different LOS of the same cloud typically have significantly different $L_\rmn{^{12}CO}$/$L_\rmn{[CII]}$ values with the exception of MC2-MHD (for both FB and noFB). The scatter is of  the order of a factor of a few in some snapshots: for instance, in MC2-HD-noFB at $t_\rmn{evol}~=$~4~Myr, $L_\rmn{^{12}CO}$/$L_\rmn{[CII]}$ is $\sim$1.5 for the LOS along the $z$-axis, but it is $\sim$4.3 along the $x$-axis. Considering $^{13}$CO instead of $^{12}$CO considerably reduces the scatter between different LOS. The same snapshot gives $L_\rmn{^{13}CO}$/$L_\rmn{[CII]} \simeq$0.4 along the $z$-axis and 0.5 along the $x$-axis. As discussed, this is due to the lower optical depth of $^{13}$CO with respect to $^{12}$CO \citep{Borchert2022}. Indeed, we would expect identical values for different LOS if the lines were completely optically thin. To a good approximation, this is the case for the snapshots which have very low CO column densities, i.e.~MC2-MHD, which is a more diffuse cloud, and MC1-HD-FB and MC2-HD-FB at \hbox{$t_\rmn{evol}$ = 4 Myr} (see Fig.~\ref{fig:abund_over_time}), where stellar feedback has dispersed most of the dense regions.

We also observe a relevant scatter of a factor of up to a few in the line ratios among different clouds for a selected H$_2$ mass fraction. This is a consequence of the different structures and properties of the clouds and does not change significantly when considering $^{12}$CO or $^{13}$CO. For instance, for $M_\rmn{H_2}/M_\rmn{H,tot} \simeq 0.3$, $L_\rmn{^{12}CO}$/$L_\rmn{[CII]}$ ranges from 2 to 6 and $L_\rmn{^{13}CO}$/$L_\rmn{[CII]}$ from 0.2 to 0.8. As we discuss in detail in the following, we do not observe a systematic relation between higher/lower line ratios (at fixed H$_2$ mass fraction) and the presence/absence of magnetic fields or stellar feedback.

Stellar feedback has an impact on the evolution of the line ratio: if no feedback is considered, the line ratios increase with $t_\rmn{evol}$: this can be seen clearly for the HD runs, and, even though less evidently, also in the MHD runs. Conversely, including stellar feedback causes a decreasing ratio over time for the same clouds. This trend is less pronounced for MC2-MHD because it is less developed and therefore the stellar feedback has a smaller impact. 

The luminosity ratios as a function of $M_\rmn{H_2}/M_\rmn{H,tot}$ exhibit an overall increasing relation for noFB runs (apart from optical thickness effects). On the other hand, for the FB clouds, there is no clear trend any more. This is a consequence of the fact that both the H$_2$ mass fraction (Fig.~\ref{fig:abund_over_time}) and the CO/[CII] luminosity ratio (Fig.~\ref{fig:line_ratio_global_fit}, left hand-side plot) decrease with $t_\rmn{evol}$, but with different slopes, making their reciprocal relation non-trivial. 

In summary, there is no clear trend of the luminosity ratios with both $t_\rmn{evol}$ and $M_\rmn{H_2}/M_\rmn{H,tot}$. This implies that it is not possible to quantitatively assess an age or evolutionary stage of a cloud by measuring a certain line ratio value. Moreover, there is a large scatter among different LOS and different clouds. We refer to Section \ref{sec:variability_line_ratio} for a further discussion about the implications of these findings and a comparisons with recent observational works.

\subsection{Analysis of single pixels}\label{sec:pixel}

\subsubsection{The intensity - column density relation}\label{sec:i_n_relation}
\begin{figure*}
    \centering
    \includegraphics[width=0.99\textwidth]{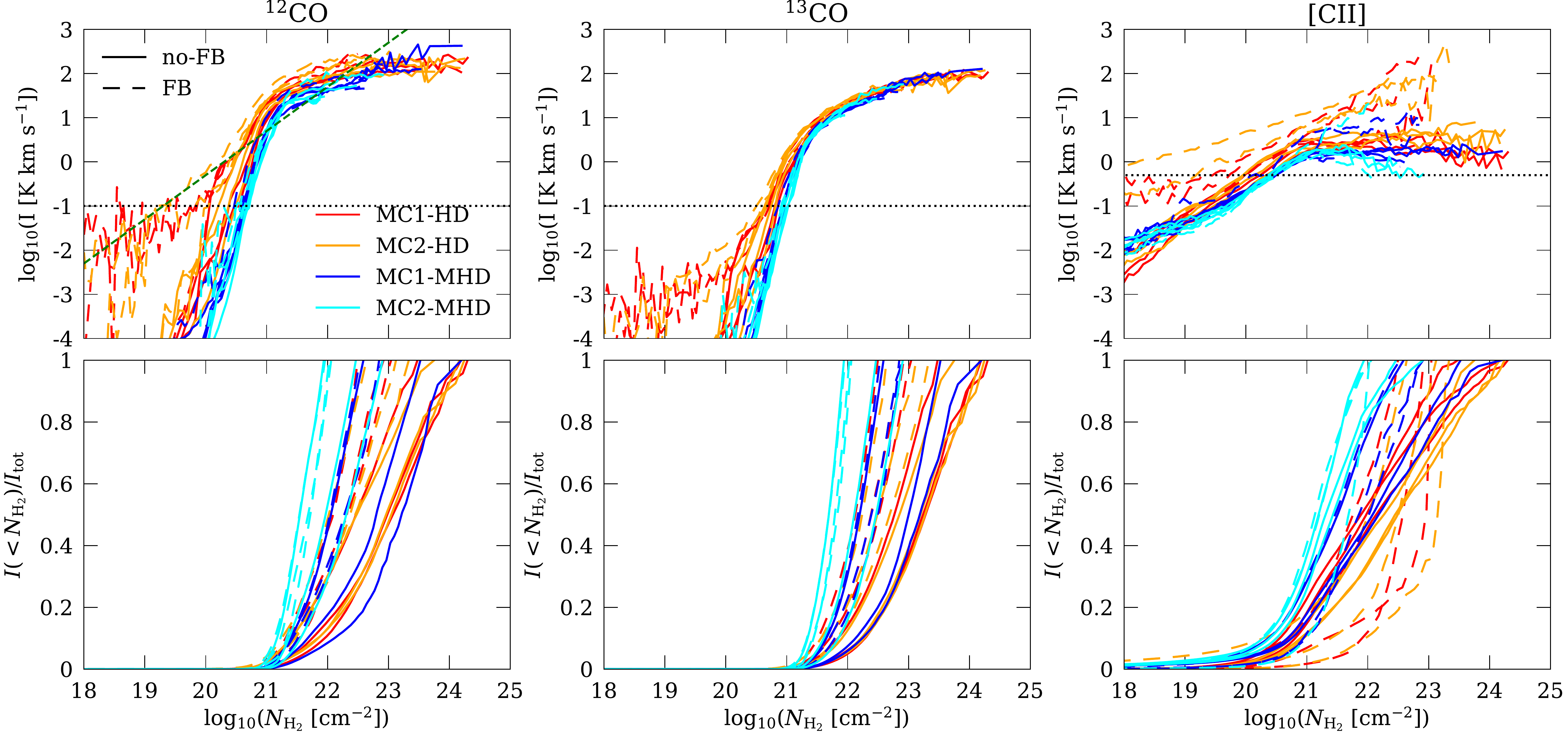}
    \caption{\textit{Top row:} $I_\rmn{^{12}CO}$, $I_\rmn{^{13}CO}$, and $I_\rmn{[CII]}$ (from left to right) as a function of the column density of H$_2$ column density see along the $z$-axis. Lines (solid for noFB runs and dashed for FB runs) represent the mean values for each $N_\rmn{H_2}$-bin. Snapshots at different $t_\rmn{evol}$, for a given cloud, are plotted with the same color. The dotted, horizontal lines represent realistic observable limits for CO and [CII], which we set to 0.1  and 0.5 K km s$^{-1}$, respectively. The dashed, green line in the top-left panel represents the $X_\rmn{CO}$ reference value given by \citet{Bolatto2013}. Optical thickness effects play a role for $N_\rmn{H_2} \gtrsim 10^{21}$ cm$^{-2}$. \textit{Bottom row:} cumulative distribution of the intensity of $^{12}$CO, $^{13}$CO, and [CII] (from left to right) as a function of a $N_\rmn{H_2}$ threshold value. Regions with $N_\rmn{H_2} < 10^{21}$ cm$^{-2}$ are associated with a negligible CO emission, but they correspond to 5 - 20 per cent of the [CII] emission. The role of feedback is evident in MC1-HD and MC2-HD, where the most of [CII] emission comes from regions with $N_\rmn{H_2} > 10^{22}$ cm$^{-2}$, corresponding to the rims of the bubbles.}\label{fig:int_vs_col_dens}
\end{figure*}

Next, we investigate -- pixel by pixel -- the relation between the intensity $I$ (in K km s$^{-1}$) of $^{12}$CO, $^{13}$CO and [CII] and the column density $N$ of $^{12}$CO, $^{13}$CO, C$^+$, and H$_2$. This is shown in Fig.~\ref{fig:int_vs_col_dens}, where in the top row $I_\rmn{^{12}CO}$ (left panel), $I_\rmn{^{13}CO}$ (middle panel), and $I_\rmn{[CII]}$ (right panel) are plotted as a function of $N_\rmn{H_2}$.  Each line represents the mean value of $I$ for a given $N_\rmn{H_2}$-bin for the selected snapshot. The bottom row shows the cumulative distribution of the intensity arising from regions with H$_2$ column densities lower or equal to the threshold of $N_\rmn{H_2}$ given on the $x$-axis. Snapshots corresponding to different $t_\rmn{evol}$ for the same cloud are plotted with the same color, and therefore are not distinguished here. We plot only the data resulting from the integration along the $z$-direction. However, we obtain qualitatively and quantitatively similar results when considering the $x$- or the $y$-direction. We note that the large scatter at very low and very high $N_\rmn{H_2}$ is due to the low number of pixels in these regimes.

The column density at which $I_{\rmn{^{12}CO}}$ and $I_\rmn{[CII]}$ become optically thick is around $N_\rmn{H_2} \simeq 10^{21}$ cm$^{-2}$ in both cases. The $I_\rmn{^{13}CO}$ values (middle panel of Fig.~\ref{fig:int_vs_col_dens}) show a similar behaviour, but the effect of optical thickness is less evident: deviations from the optically thin behaviour occur at $N_\rmn{H_2} \gtrsim 10^{21}$ cm$^{-2}$, but the change of slope is less pronounced than in the $^{12}$CO case. The kink in the $I_\rmn{[CII]}$ - $N_\rmn{H_2}$ relation does not appear to happen for most of the FB runs. We attribute it to the fact that the regions, where $N_\rmn{H_2} > 10^{21}$ cm$^{-2},$ are essentially the rims of the expanding bubbles (see Fig.~\ref{fig:col_dens_picture}). In these regions [CII] emission is enhanced by the effect of stellar radiation (see Section~\ref{sec:bubbles}) and -- even if optically thick -- is thus larger than the intensity of a few K km s$^{-1}$ coming from non-irradiated optically thick regions. 

The values of $N_\rmn{H_2}$ at which $I_\rmn{CO}$ and $I_\rmn{[CII]}$ become optically thick are in a good agreement with, for instance, the simulations of \citet{Bisbas2021}, in particular regarding the $I_\rmn{^{12}CO}$ - $N_\rmn{H_2}$ relation, whereas the $I_\rmn{[CII]}$ - $N_\rmn{H_2}$ is more dependent on the environmental conditions of the simulated clouds. Moreover, \citet{Beuther2014} find an $I_\rmn{[CII]}$ - $N_{H_2}$ relation in the G48.66 cloud in good agreement with our curve, whereas a study of the Perseus Giant Molecular Cloud by \citet{Hall2020} shows an overall lower $I_\rmn{[CII]}$ for given $N_\rmn{H_2}$. Finally, comparing the $I_\rmn{^{12}CO} - N_\rmn{H_2}$ with the typical $X_\rmn{CO}$ of $N_\rmn{H_2}/I_\rmn{^{12}CO} = 2 \times 10^{20}$ cm$^{-2}$ K$^{-1}$ km$^{-1}$ \citep[][green line]{Bolatto2013} shows that a linear relation between the two quantities does not hold on local scales, as already pointed out in \citet{Seifried2020}.
 
In Fig.~\ref{fig:lum_over_time} we have already shown that the [CII] luminosity is enhanced by stellar feedback. Next, we analyse the column density regimes where the majority of luminosity comes from. For $N_\rmn{H_2} \geq 10^{22}$ cm$^{-2}$ we report typical $I_\rmn{[CII]}$ values of 1 - 10 K km s$^{-1}$ for noFB runs and up to a few $\sim 10^2$ K km s$^{-1}$ for feedback runs (right panel of Fig.~\ref{fig:int_vs_col_dens}). Stellar feedback increases the [CII] excitation temperature, which causes a stronger emission for a given density. In general, areas with the highest $N_\rmn{C^+}$ are those closer to the star-forming regions (see Fig.~\ref{fig:col_dens_picture}). Therefore, this is also the regime where the difference between FB and noFB runs are most evident. In addition, for the HD clouds (red and orange lines) at late evolutionary stages, stellar feedback affects even larger parts of the clouds, such that the enhancement in $I_\rmn{[CII]}$ is visible also in lower column density regimes. 

The cumulative distribution of the intensity (bottom row of Fig.~\ref{fig:int_vs_col_dens}) reveals the column density range from which the line emission comes from. We have no CO emitting regions with $N_\rmn{H_2} < 10^{21}$ cm$^{-2}$, whereas 5 - 20 per cent of the [CII] emission comes from these areas. Despite this, we still argue that a significant fraction of [CII] comes from the atomic phase \citep{Franeck2018} as also pixels with high $N_\rmn{H_2}$ still do have a significant amount of atomic hydrogen mixed in (and thus a high $N_\rmn{H}$) \citep{Seifried2022}. In general, the lines in the plot corresponding to the different clouds differ as each cloud has a different maximum $N_\rmn{H_2}$ value. The role of feedback on the [CII] emission is particularly evident: a large part of the emission comes from the rims of the bubbles, which typically regions with $N_\rmn{H_2} \gtrsim 10^{22}$ cm$^{-2}$. This is reflected in the steep ascent of the corresponding curves at \mbox{$N_\rmn{H_2} > 10^{22}$ cm$^{-2}$} for MC1-HD-FB and MC2-HD-FB, the clouds which are most affected by stellar feedback.

\begin{figure}
    \centering
    \includegraphics[width=0.99\columnwidth]{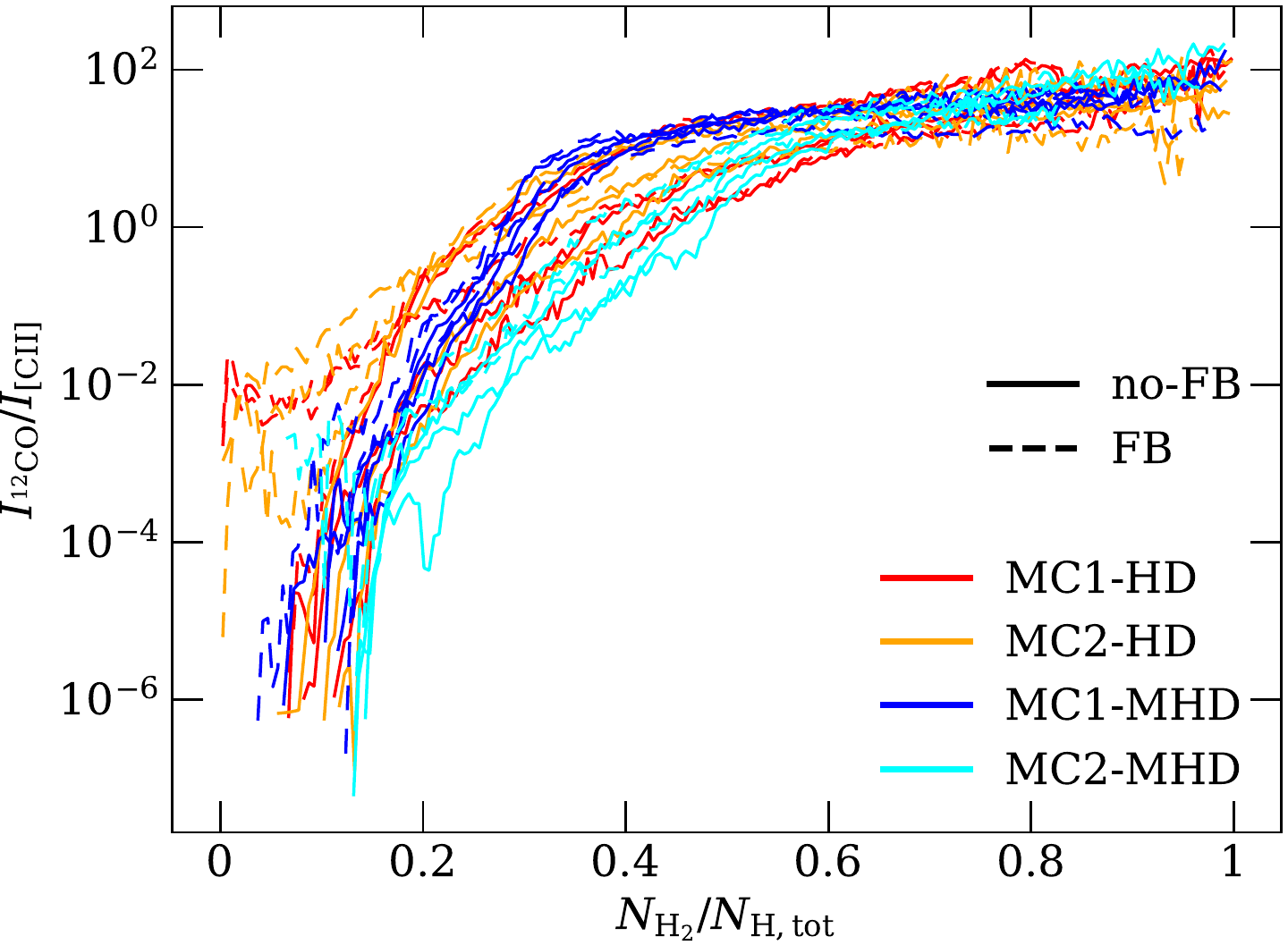}
    \caption{CO/[CII] intensity ratio plotted as a function of the H$_2$ mass fraction $N_\rmn{H_2}/N_\rmn{H,tot}$. We plot the snapshots at different $t_\rmn{evol}$ with the same color. The line ratio increases with $N_\rmn{H_2}/N_\rmn{H,tot}$, although there is a large scatter, which is particularly relevant at high $N_\rmn{H_2}/N_\rmn{H,tot}$ as the relation is much shallower in this regime than at lower mass fraction. This prevents the usage of $I_\rmn{^12CO}/I_\rmn{[CII]}$ to determine $N_\rmn{H_2}/N_\rmn{H,tot}$.} 
    \label{fig:line_ratio_pixel}
\end{figure}

\begin{figure}
    \centering
    \includegraphics[width=0.99\columnwidth]{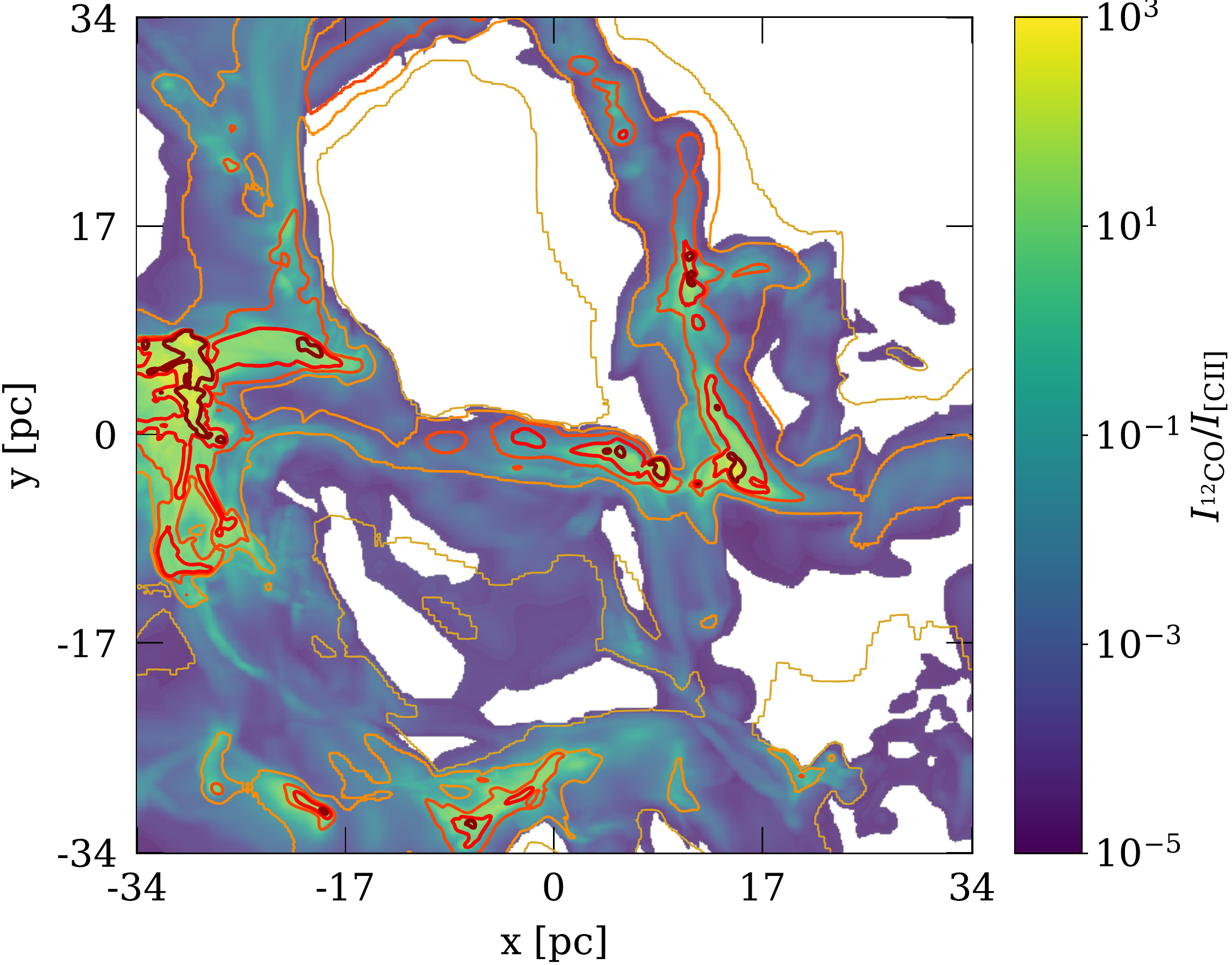}
    \caption{Map of $I_\rmn{^{12}CO}/I_\rmn{[CII]}$ for MC1-HD at $t_\rmn{evol} = 4$ Myr. Contour lines indicate an H$_2$ mass fraction $N_\rmn{H_2}/N_\rmn{H,tot}$ of 0.1, 0.3, 0.5, 0.7 and 0.9, respectively. We note a general correspondence between higher intensity ratio and higher $N_\rmn{H_2}/N_\rmn{H,tot}$ values. There are, however, significant differences in the line ratio for a given $N_\rmn{H_2}/N_\rmn{H,tot}$ value, especially in moderate $N_\rmn{H_2}/N_\rmn{H,tot}$ regimes.}
    \label{fig:int_ratio_map}
\end{figure}

\subsubsection{$I_\rmn{CO}$ / $I_\rmn{[CII]}$ and H$_2$ mass fraction}\label{sec:intensity_ratio}

In Section~\ref{sec:line_ratio} we show that there is no clear relation between the CO/[CII] luminosity ratio and the H$_2$ mass fraction. Now, we consider the intensity ratio $I_\rmn{^{12}CO}/I_\rmn{[CII]}$ as a function of $N_\rmn{H_2} / N_\rmn{H,tot}$ for each pixel of the maps. We need to be sure that we consider the same portion of the cloud for both, CO and C$^+$. For this purpose, we loop over the 201 velocity channels of the maps and indicate with $\{k \}$ the set of channels for which both specific intensities, $I_{v,k,\rmn{^{12}CO}}$ and $I_{v,k,\rmn{[CII]}}$, are above the specific intensity of the Cosmic Microwave Background, corresponding to 0.84 and $2.69 \times 10^{-13}$ K for $^{12}$CO and [CII], respectively. (Eq.~\ref{eq:T_cmb}). We define 
\begin{equation}
    I_\rmn{^{12}CO} = \sum_k I_{v,k,\rmn{^{12}CO}} \Delta v \; ,
\end{equation}
where $\Delta v$ is the width of a velocity channel, $I_\rmn{[CII]}$ is obtained analogously. The result is shown in Fig.~\ref{fig:line_ratio_pixel}, where the mean values of the distribution of $I_\mathrm{CO} / I_\rmn{[CII]}$ for a given $N_\rmn{H_2}$ are shown using again the same color for snapshots referring to different $t_\rmn{evol}$. We show here the data referring to the LOS along the $z$-axis, but we obtain analogous results for the integration along the other LOS. The $I_\rmn{^{12}CO} / I_\rmn{[CII]}$ ratio increases with increasing $N_\rmn{H_2} / N_\rmn{H, tot}$. However, different clouds and snapshots show significantly different line ratios for given $N_\rmn{H_2} / N_\rmn{H,tot}$ with a typical scatter of up to two orders of magnitude. At very low values of $N_\rmn{H_2} / N_\rmn{H,tot}$ the scatter is even larger due to the low statistics. Furthermore, the presence of CO-dark and CO-bright pixels for the same $N_\rmn{H_2}$ \citep[as shown in detail in][see their figure~8]{Seifried2020} also contributes to enlarge such scatter. 

Hence, the variability for a given $N_\rmn{H_2}/N_\rmn{H, tot}$ value is so large that the ratio $I_\rmn{^{12}CO}/I_\rmn{[CII]}$ cannot be reliably used to determine $N_\rmn{H_2}/N_\rmn{H, tot}$. This is also shown in the example given in Fig.~\ref{fig:int_ratio_map}, where we show a map of $I_\rmn{^{12}CO}/I_\rmn{[CII]}$ for MC1-HD-noFB at $t_\rmn{evol} = 4$ Myr. We overplot isocontour lines corresponding to H$_2$ mass fractions of 0.1, 0.3, 0.5, 0.7, and 0.9. There is a general correspondence between high line ratios and high H$_2$ mass fractions, but there are still significant variations in the line ratio within regions of similar H$_2$ mass fraction, especially for mass fraction regimes between 0.3 and 0.7.

Our results for the \textit{pixel-by-pixel} approach are thus similar to that for the global luminosity ratio shown in Fig.~\ref{fig:line_ratio_global_fit}, which also does not allow for a determination of the global H$_2$ mass fraction. However, we cannot directly compare the relation between $I_\rmn{^{12}CO} / I_\rmn{[CII]}$ and $N_\rmn{H_2} / N_\rmn{H, tot}$ with the one between $L_\rmn{^{12}CO} / L_\rmn{[CII]}$ and $M_\rmn{H_2} / M_\rmn{H, tot}$. For the pixel-by-pixel approach we also find H$_2$ column density fractions close to 0 and 1 and corresponding $I_\rmn{^{12}CO} / I_\rmn{[CII]}$ values spanning 8 orders of magnitude. On the other hand,when analysing $L_\rmn{^{12}CO} / L_\rmn{[CII]}$, we average over the entire cloud and, as a consequence, both the mass fraction and the luminosity ratio span over a considerably lower range.

\section{Discussion}\label{sec:discussion}

\subsection{Intrinsic variability of line ratios}\label{sec:variability_line_ratio}

The $^{12}$CO/[CII] and $^{13}$CO/[CII] line ratios shown in Fig.~\ref{fig:line_ratio_global_fit} are characterized by a large dispersion due to the difference in the structure and evolutionary stage of the clouds themselves only, but not to different environments or external factors. In fact, all clouds form in a portion of a galactic disc with the same CRIR, $G_0$, and metallicity and with turbulence driven by supernovae. Furthermore, all MHD runs have the same initial magnetic field strength. Despite that, $L_\rmn{^{12}CO}/L_\rmn{[CII]}$ varies by up to a factor of 5 for a given $M_\rmn{H_2} / M_\rmn{H,tot}$. The same applies for $^{13}$CO, although the scatter among different LOS for the same snapshot is reduced because of the reduced optical depth.

This scatter is also found in other works. For instance, \citet{Rollig2006} use the luminosity ratio to assess the environmental conditions like the cloud metallicity, density, and FUV field intensity\footnote{In order to compare values from observational works with our simulations, it might be necessary to convert the intensity from \mbox{K km s$^{-1}$} to \mbox{erg s$^{-1}$ cm$^{-2}$} and recalculate the luminosity and luminosity ratio (see Fig.~\ref{fig:line_ratio_erg}). Note that we plot [CII]/CO there, whereas Fig.~\ref{fig:line_ratio_global_fit} shows CO/[CII].}. Their models also exhibit large difference in the line ratio up to a factor of a few, even when leaving environmental conditions like the metallicity, FUV, and cloud density unchanged. Furthermore, also observational results at similar metallicities  quoted in their work, e.g. for the LMC and 30 Doradus, show a similar scatter of 5 - 10 in the [CII]/$^{12}$CO line ratio. In summary, as already stated by \citet{Rollig2006}, we do not recommend to use $L_\rmn{^{12}CO}/L_\rmn{[CII]}$ to infer physical properties of the clouds.

Furthermore, \citet{Madden2020} analyse the $^{12}$CO and [CII] emission in a variety of environments. For normal galaxies and galactic star-forming regions, they find $L_\rmn{[CII]}/L_\rmn{^{12}CO} \approx 4000$, with a large scatter covering values from 300 to $25\,000$. Our simulation results are thus in good agreement with their findings, although they consider a much larger variety of environments.

Recently, \citet{Hall2020} analysed the $^{12}$CO (1 $\rightarrow$ 0) and [CII] emission from two regions of the Perseus Giant Molecular Cloud. As their observations refer to a resolved portion of a cloud, this corresponds to our \textit{pixel-by-pixel} analysis shown in Fig.~\ref{fig:line_ratio_pixel}. In general, we find that their reported values of $I_\rmn{^{12}CO}/I_\rmn{[CII]}$ of 2 -- 100 agree well with ours. They also show that the highest values of the line ratio are reached where H$_2$ (obtained via a comparison of the optical depth, obtained from dust continuum measurements, and the HI column density, see their Eq.~1) is more abundant, which is in agreement with our results. 

Finally, \citet{Bisbas2021} analysed the line emission of several species from two different, simulated clouds with  different environmental parameters. For comparable CRIR, $G_0$, and metallicity, they find $^{12}$CO/[CII] line ratios larger  by up to one order of magnitude compared to our work. We tentatively attribute this difference to two main factors. First, our clouds are somewhat more diffuse than the clouds used by \citet{Bisbas2021} (priv. communication). Indeed, when going to later evolutionary states, i.e.~denser clouds, our line ratios increase (Fig.~\ref{fig:line_ratio_global_fit}). Second, their work assumes chemical equilibrium, while we use non-equilibrium chemistry, a difference whose effects we will discuss in detail in the following.

\subsection{Equilibrium vs.~Non-equilibrium chemistry}
\label{sec:equilibrium}

\begin{figure*}
    \centering
    \includegraphics[width=0.99\textwidth]{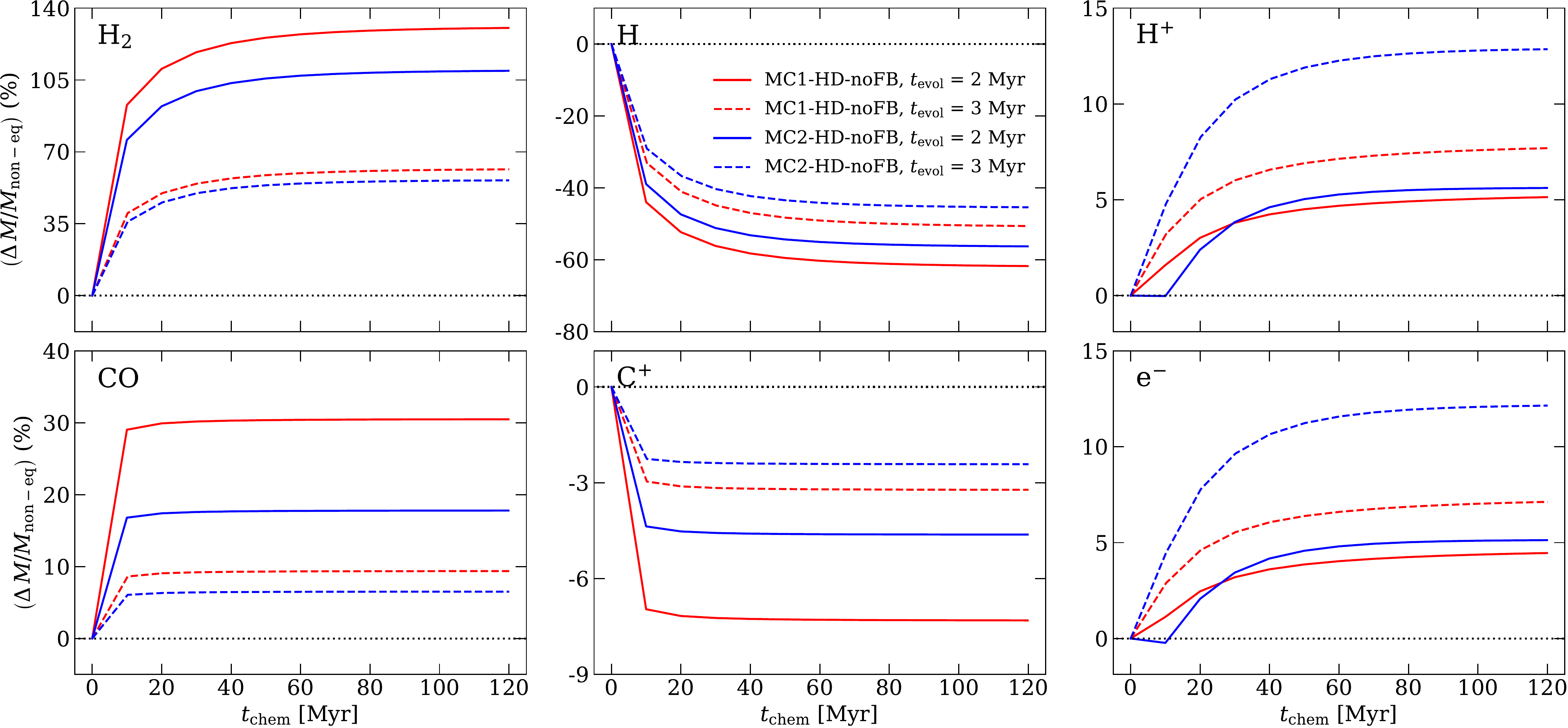}
    \caption{Relative mass changes of H$_2$, H, H$^+$, CO, C$^+$, and e$^-$ (from top left to bottom right) for a selection of snapshots, obtained by freezing the evolution of the physics in the simulation and only evolving the chemistry for a time $t_\rmn{chem}$. CO and C$^+$ reach equilibrium after $\sim 10$ Myr, whereas H and H$_2$ reach equilibrium at $\gtrsim 40$ Myr. The mass variations at equilibrium are particularly important for H (decreasing by up to 60 per cent) and for H$_2$ (increasing by up to 120 per cent), making the assumption of equilibrium for these species questionable. Changes for carbon-bearing species, H$^+$, and e$^-$ are smaller than for H and H$_2$.} 
    \label{fig:mass_equilibrium}
\end{figure*}
A large number of MC simulation works post-process their results to obtain chemical abundances by assuming that the chemical state is in equilibrium \citep[see e.g.][]{Gong2018, Li2018, Keating2020,Bisbas2021}. As an example, \citet{Gong2020} study the $X_\rmn{CO}$ factor for a wide range of environments. They evolve the chemical network for \hbox{50 Myr}, i.e. equilibrium is roughly reached at that point \citep{Joshi2019}, before analysing synthetic CO observations and the $X_\rmn{CO}$ factor. However, the presence of phenomena like e.g.~the turbulent mixing \citep[see e.g.][]{Glover2007c, Valdivia2016, Seifried2017} suggests that this approach can determine inaccurate estimations at least for hydrogen and directly related quantities like e.g.~$X_\rmn{CO}$.

Here, we aim to assess how much the assumption of equilibrium chemistry affects synthetic emission maps. In order to do so, we first select the snapshots of MC1-HD-noFB and MC2-HD-noFB at $t_\rmn{evol}~=~2$ and 3~Myr and only evolve the chemistry with the NL97 network for additional 120~Myr while the hydrodynamical state (total gas density, etc.) remains frozen. In the following, $t_\rmn{chem}$ refers to the time for which the chemistry of the snapshot was evolved. We also define $M_\rmn{no-eq} = M(t_\rmn{chem} = 0)$ and $L_\rmn{no-eq} = L(t_\rmn{chem} = 0)$.

Fig.~\ref{fig:mass_equilibrium} shows the evolution of H$_2$, H, H$^+$, CO, C$^+$, and e$^-$ as a function of $t_\rmn{chem}$. The CO and H$_2$ masses (left column) increase with $t_\rmn{chem}$ by up to 30 and 120 per cent, respectively, in rough agreement with e.g.~\citet{Gong2018}. The masses of H$^+$ and e$^-$ (right column) also increase over time but the increase is less pronounced (between 3 and 12 per cent). The masses of C$^+$ and H decrease with $t_\rmn{chem}$ by 7 and 60 per cent, respectively. In summary, these results once again confirm that the assumption of chemical equilibrium is -- in particular for hydrogen-bearing species -- questionable \citep{Glover2007c, Valdivia2016, Seifried2017, Seifried2022, Chiayu2021}.

Furthermore, we find that CO and C$^+$ reach equilibrium at $t_\rmn{chem} \sim$ 10 Myr, whereas H$_2$, H, and H$^+$ reach it after $\gtrsim$ 40 Myr. In all cases, however, we can assume that at $t_\rmn{chem}$~=~50~Myr (used in the following) equilibrium is roughly reached, as the relative changes with respect to later times are $\lesssim~5$ per cent. 

\begin{figure}
    \centering
    \includegraphics[width=0.99\columnwidth]{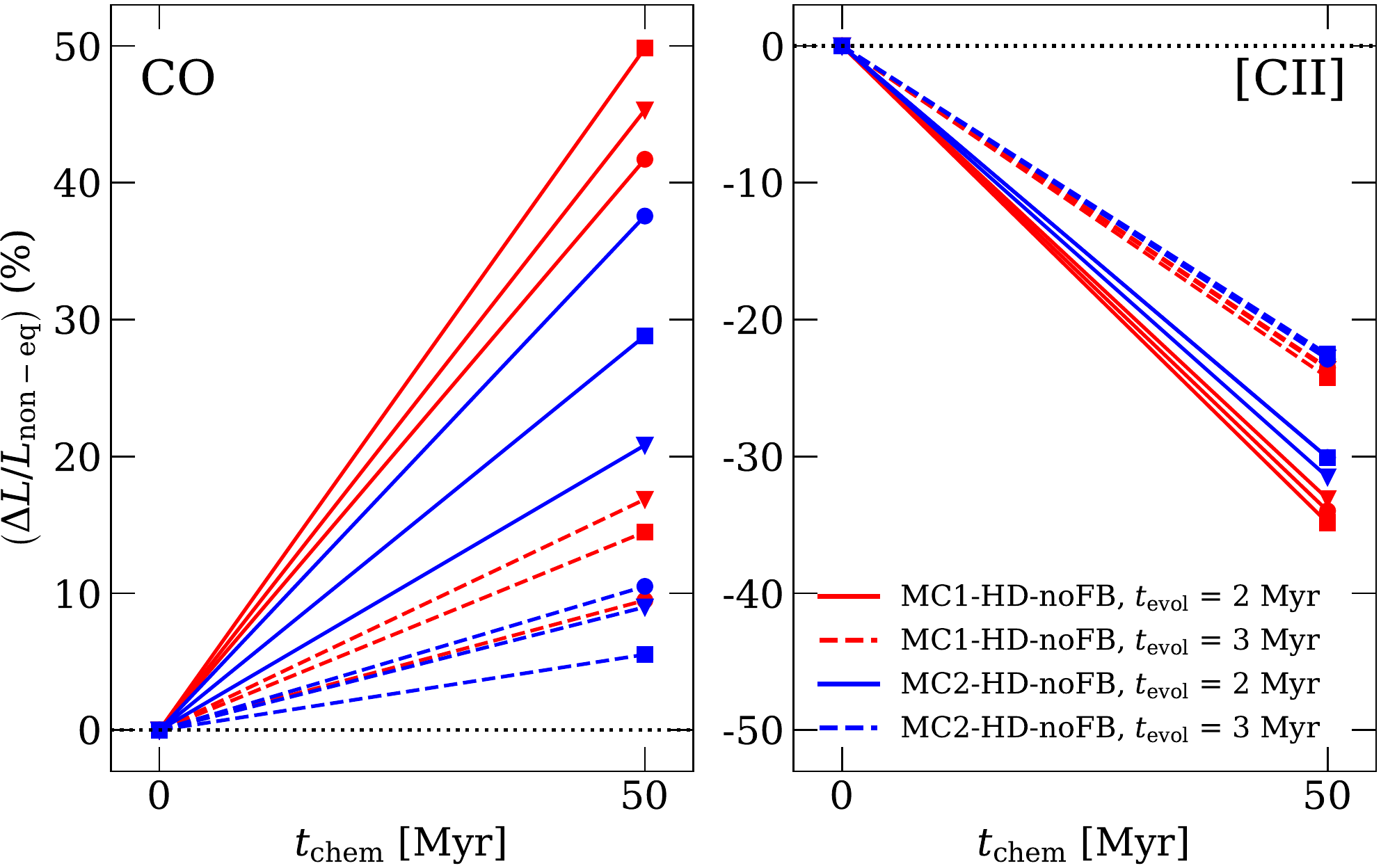}
    \caption{Relative luminosity changes of $L_\rmn{CO}$ (left) and $L_\rmn{[CII]}$ (right) between $t_\rmn{chem} = 0$ and $t_\rmn{chem} = 50$ Myr for the same selection of snapshot as in Fig.~\ref{fig:mass_equilibrium}. The three different LOS are indicated with different markers. We note that the the luminosity decreases for [CII] up to $\sim 30$ per cent, while the corresponding mass decreases of $\sim 7$ per cent. The changes in the CO luminosity are comparable with the changes in the CO mass.} 
    \label{fig:intensity_equilibrium}
\end{figure}

The significant changes of the chemical abundances by up to 120 per cent also affect the synthetic emission maps. In Fig.~\ref{fig:intensity_equilibrium} we show the difference in total luminosity between the equilibrium (defined here as the state at $t_\rmn{chem} = 50$ Myr) and non-equilibrium state ($t_\rmn{chem} = 0$) for $^{12}$CO and [CII]. The luminosity of CO increases in equilibrium by up to $50$ per cent with respect to the non-equilibrium case. This increase is only marginally larger than the respective increase of the CO mass (up to $30$ per cent), indicating that the increase in $M_\rmn{CO}$ in chemical equilibrium is responsible for this luminosity change. Furthermore, due to the different gain in $M_\rmn{H_2}$ (top left panel of Fig.~\ref{fig:mass_equilibrium}) and $L_\rmn{CO}$, the value of $X_\rmn{CO}$ determined for the chemical equilibrium case is about 50 per cent larger than the actual value for the non-equilibrium state. However, this deviation is within the typical scatter of $X_\rmn{CO}$ of a factor of a few found here (see Section~\ref{sec:x_co}) as well as in \citet{Gong2020} using equilibrium chemistry. Hence, differences caused by the equilibrium approach can hardly be assessed by comparing the values for $X_\rmn{CO}$ obtained in both works.

On the other hand, the change in [CII] luminosity (right panel of Fig.~\ref{fig:intensity_equilibrium}) is considerably larger than the corresponding change in mass (as shown in Fig.~\ref{fig:mass_equilibrium}), e.g.~for MC1-HD-noFB at 2~Myr for one LOS we have $\Delta L/L_\rmn{non-eq} \simeq - 30$ per cent and $\Delta M/M_\rmn{non-eq} \simeq - 7$ per cent.

One element contributing to explain the changes in mass and luminosity are the collisional partners, which in our case are H$_2$, H, and electrons. We find that the H$_2$ abundance increases with $t_\rmn{chem}$, whereas H decreases. Although the electron abundance increases for the equilibrium case and the C$^+$ $-$ e$^-$ de-excitation rates rates are in general larger than those of H and H$_2$, this does not lead to an increase in $L_\rmn{[CII]}$. We attribute this to the fact that the relative change of the electron abundance is significantly lower ($\lesssim 12$ per cent) than for the other two collisional partners. Additionally, there is little C$^+$ in the low-density/high-temperature regime where the e$^-$ collisional rate is high. Hence, as the [CII] emission is dominated by atomic gas \citep{Franeck2018}, the drop in H mass is mainly responsible for the drop in $L_\rmn{[CII]}$.

The impact of the collisional partners affecting $L_\rmn{[CII]}$ can also be expressed by the excitation temperature (see Fig.~\ref{fig:exc_temperature} in the appendix). We find that $T_\rmn{ex}$ is overall lower for the equilibrium case. This contributes to explain the larger decrease in luminosity than in mass when moving to equilibrium for C$^+$. 

Another factor explaining why the relative changes of mass and luminosity for C$^+$ do not directly correlate is connected to the detailed distribution of C$^+$ in the density -- temperature phase space. The majority of C$^+$ mass is contained in the Warm Neutral Medium (WNM), where the C$^+$ abundance is already quite close to chemical equilibrium. Thus, evolving the chemistry to equilibrium does not imply a major change in this region and then the overall change in $M_\rmn{C^+}$ is rather moderate. However, for observations towards MCs, the C$^+$ in the WNM -- despite existing in this environment -- contributes only little to the total [CII] luminosity. Rather, most of the [CII] luminosity from MCs comes from the Cold Neutral Medium (CNM) \citep{Franeck2018}. In the CNM, however, the C$^+$ abundance is further away from equilibrium, hence evolving the chemistry to equilibrium produces a significant change in the C$^+$ mass in this region, and in consequence on the total [CII] luminosity. The total change in C$^{+}$ mass (being dominated by the WNM) is, however, minor.

We emphasise that the values for \hbox{$t_\rmn{evol} = 2$ Myr} (solid lines in Figs.~\ref{fig:mass_equilibrium} and~\ref{fig:intensity_equilibrium}) change more, both in terms of mass and luminosity, than the values for \hbox{$t_\rmn{evol} = 3$ Myr} (dashed lines). Hence, early evolutionary stages appear to be further away from a chemical equilibrium state, as the overall densities are still lower and thus, the  chemical timescales longer. As a result, we argue that the chemical post-processing of MC simulations up to equilibrium, in particular at an early evolutionary state, is questionable and should be considered with great caution.

Given the luminosity changes shown in Fig.~\ref{fig:intensity_equilibrium}, the values of $L_\rmn{CO}/L_\rmn{[CII]}$ (not shown) increase by up to 100 per cent for the equilibrium case with respect to the non-equilibrium case. Assuming chemical equilibrium can therefore lead to a relative error of up to a factor of $\sim2$ when calculating such line ratios. As pointed out before, this error is generally larger at early evolutionary stages of the clouds. This effect can thus contribute to the differences seen in line ratios when compared to e.g. the work of \citet{Bisbas2021}.

\begin{figure}
    \centering
    \includegraphics[width=0.99\columnwidth]{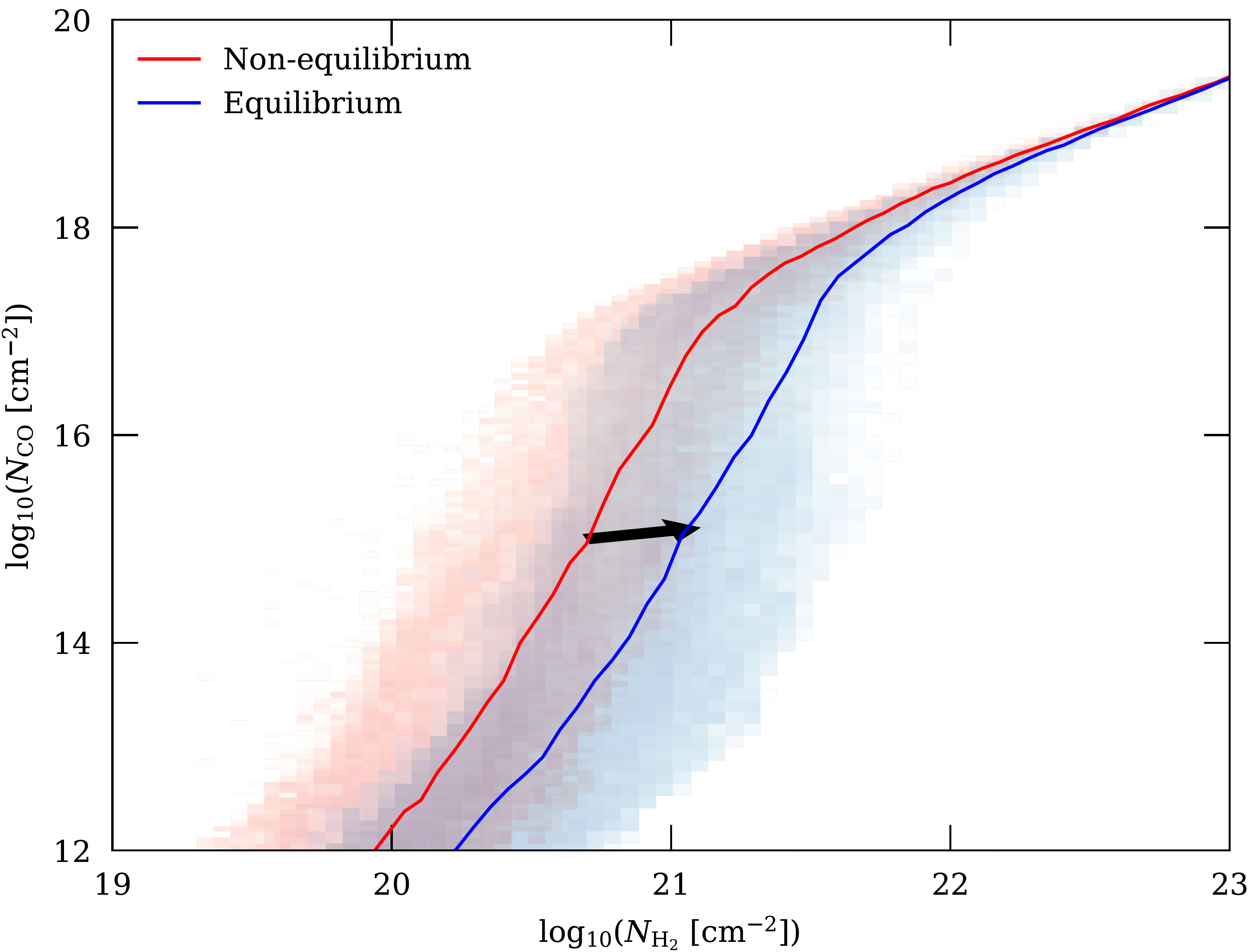}
    \caption{$N_\rmn{CO}$ as a function of $N_\rmn{H_2}$ for MC2-HD at $t_\rmn{evol} = 2$ Myr, considering the chemical state in non-equilibrium (red) and at equilibrium (blue). Shaded areas represent the 2D-PDFs and solid lines represent the mean values. The change is due to the larger increase in the H$_2$ mass than in the CO mass when moving from non-equilibrium to equilibrium chemistry (120 per cent vs. 30 per cent for the considered snapshot). The black arrow qualitatively indicates the change of the abundances for this transition.}
    \label{fig:co_vs_h2_equilibrium}
\end{figure}

Finally, we investigate the change in the relation between $N_\rmn{H_2}$ and $N_\rmn{CO}$ assuming chemical equilibrium. In Fig.~\ref{fig:co_vs_h2_equilibrium} we show an example for MC2-HD-noFB at $t_\rmn{evol}~=~2$~Myr. The red-shaded area represents the 2D-PDF for the original, non-equilibrium snapshot, whereas the blue-shaded area represents the equilibrium case. The two lines indicate the mean values. We observe a shift towards higher $N_\rmn{H_2}$ for a given $N_\rmn{CO}$ for chemical equilibrium, which we mainly attribute to the more pronounced increase in $M_\rmn{H_2}$ than in $M_\rmn{CO}$ in case of chemical equilibrium (see Fig.~\ref{fig:mass_equilibrium}). We note that this is in excellent agreement with results of \citet{Chiayu2021}, who find a similar difference in mass changes for H$_2$ and CO concerning equilibrium and non-equilibrium states.

To summarize, we consider it as crucial to use non-equilibrium chemistry to simulate the H/H$_2$ content of MCs, as cloud evolution and molecule formation go hand in hand. Because of this, using chemical equilibrium for simulated MCs should be considered with great caution, in particular at early evolutionary stages, as it can significantly effect both the masses and luminosities of the various species.

\section{Conclusions}\label{sec:conclusions}

We present an analysis of the abundance and luminosity of $^{12}$CO, $^{13}$CO and C$^+$ for 8 simulated MCs within the SILCC-Zoom project \citep{Seifried2017}, in which the chemical network is evolved \textit{on-the-fly}. In particular, we investigate two clouds with and two without magnetic fields under solar neighborhood conditions at different evolutionary stages. For each simulation we consider a reference case without stellar feedback and one including radiative feedback in the form of ionizing radiation by massive stars. For this purpose, we have developed a novel post-processing routine (based on \textsc{Cloudy}) to account for higher ionization states of carbon. We show that this post-processing is essential to obtain reliable [CII] emission maps in feedback-dominated regions.

Our conclusions can be summarised as follows:

\begin{itemize}

    \item The [CII] emission maps of the runs with radiative feedback show expanding HII regions/bubbles, where carbon is largely in form of C$^{2+}$ and thus devoid of [CII] emission inside, but with significant emission at the rims. This is in good agreement %around the star-forming regions,
    with recent [CII] surveys.
    
    \item We estimate that radiative feedback increases the [CII] luminosity by $\sim 50 - 85$ per cent compared to the non-feedback case due to an enhancement of the excitation temperature. The CO luminosity decreases by up to a factor of 3 at late evolutionary stages of the clouds due to the dispersal of dense regions. 
    
    \item The line luminosity ratios $L_\rmn{^{12}CO}/L_\rmn{[CII]}$ and $L_\rmn{^{13}CO}/L_\rmn{[CII]}$, integrated over the entire maps, show an increase with increasing H$_2$ mass fraction in noFB runs, but no clear relation in FB runs. We obtain values for $L_\rmn{^{12}CO}/L_\rmn{[CII]}$ from 1 to 6 and for $L_\rmn{^{13}CO}/L_\rmn{[CII]}$ from 0.1 to 1.1. We argue that due to the large spread, these line ratios \textit{cannot} be used as a reliable tracer of the cloud's H$_2$ mass fraction. Similarly, this spread makes it difficult to use them to assess environmental parameters like the CRIR, the IRSF, or the metallicity, which we kept fixed in our simulations.
   
    \item A pixel-by-pixel analysis of  $I_\rmn{^{12}CO} / I_\rmn{[CII]}$ as a function of $N_\rmn{H_2} / N_\rmn{H,\,tot}$ shows an increase of the ratio with $N_\rmn{H_2} / N_\rmn{H,\,tot}$. However, as for the total luminosity ratio, also here the scatter is so significant that $I_\rmn{^{12}CO} / I_\rmn{[CII]}$ cannot reliably be used to predict the mass fraction of H$_2$ along the LOS. 

    \item Evolving the chemistry to equilibrium as done in various works results in significant differences in terms of species abundance with respect to a self-consistent non-equilibrium approach used \textit{on-the-fly} during the simulation. Hence, in particular for early evolutionary stages an equilibrium approach is questionable. We find that for the equilibrium case, the H$_2$ mass is increased and the H mass is decreased by up to a factor of about~2. Other species abundances such as CO, C$^+$ and electrons change by a few 10 per cent.
    
    \item Assuming chemical equilibrium also affects the inferred luminosities of CO and [CII], with relative changes of up to $+50$ and $-30$ per cent, respectively. These luminosity changes cause an overestimate of the $L_\rmn{CO}/L_\rmn{[CII]}$ line ratios by up to $100$ per cent if equilibrium chemistry is assumed. Similarly, the $X_\rmn{CO}$ factor would be overestimated by up to 50 per cent in this case.

    \item In general, the $X_\rmn{CO}$ factor ranges between 0.5 and \mbox{4.5 $\times$ 10$^{20}$ cm$^{-2}$ K$^{-1}$ km$^{-1}$ s}, showing no clear trend with respect to time evolution or the H$_2$ mass fraction. Feedback runs in general have a lower $X_\rmn{CO}$ than the corresponding non-feedback runs. The similarly defined $X_\rmn{[CII]}$ factor ranges between 0.5 and  \hbox{12 $\times$ 10$^{20}$ cm$^{-2}$ K$^{-1}$ km$^{-1}$ s}, also not showing a clear trend with evolutionary time or H$_2$ mass fraction.
    
\end{itemize}

In summary, we show that it is crucial to take into account the effects (i) of stellar radiation in further ionizing C$^+$ within HII regions, and (ii) an \textit{on-the-fly}, non-equilibrium chemistry treatment to accurately model CO and [CII] line emission in simulated MCs. 
We thus strongly suggest to consider both effects for future and more detailed comparisons with observations (e.g.~Ebagezio et al., in prep.).

%%%%%%%%%%%%%%%%%%%%%%%%%%%%%%%%%%%%%%%%%%%%%%%%%%
\section*{Acknowledgements}
SW and PCN gratefully acknowledge the European Research Council under the European Community's Framework Programme FP8 via the ERC Starting Grant RADFEEDBACK (project number 679852). SE, DS, SW and PCN further thank the Deutsche Forschungsgemeinschaft (DFG) for funding through the SFB~956 ''The conditions and impact of star formation'' (sub-projects C5 and C6).
TN acknowledges support from the DFG under Germany’s Excellence Strategy - EXC-2094 - 390783311 from the DFG Cluster of Excellence ''ORIGINS''.
The software used in this work was in part developed by the DOE NNSA-ASC OASCR Flash Center at the University of Chicago. We particularly thank the Regional Computing Center Cologne for providing the computational facilities for this project by hosting our supercomputing cluster "Odin".



\bibliographystyle{mnras}
\bibliography{biblio} % if your bibtex file is called example.bib

%\begin{thebibliography}{99}
%\end{thebibliography}

%%%%%%%%%%%%%%%%%%%%%%%%%%%%%%%%%%%%%%%%%%%%%%%%%%

%%%%%%%%%%%%%%%%% APPENDICES %%%%%%%%%%%%%%%%%%%%%

\appendix

\section{Supplementary figures}

\begin{figure*}
    \centering
    \includegraphics[width=0.9\textwidth]{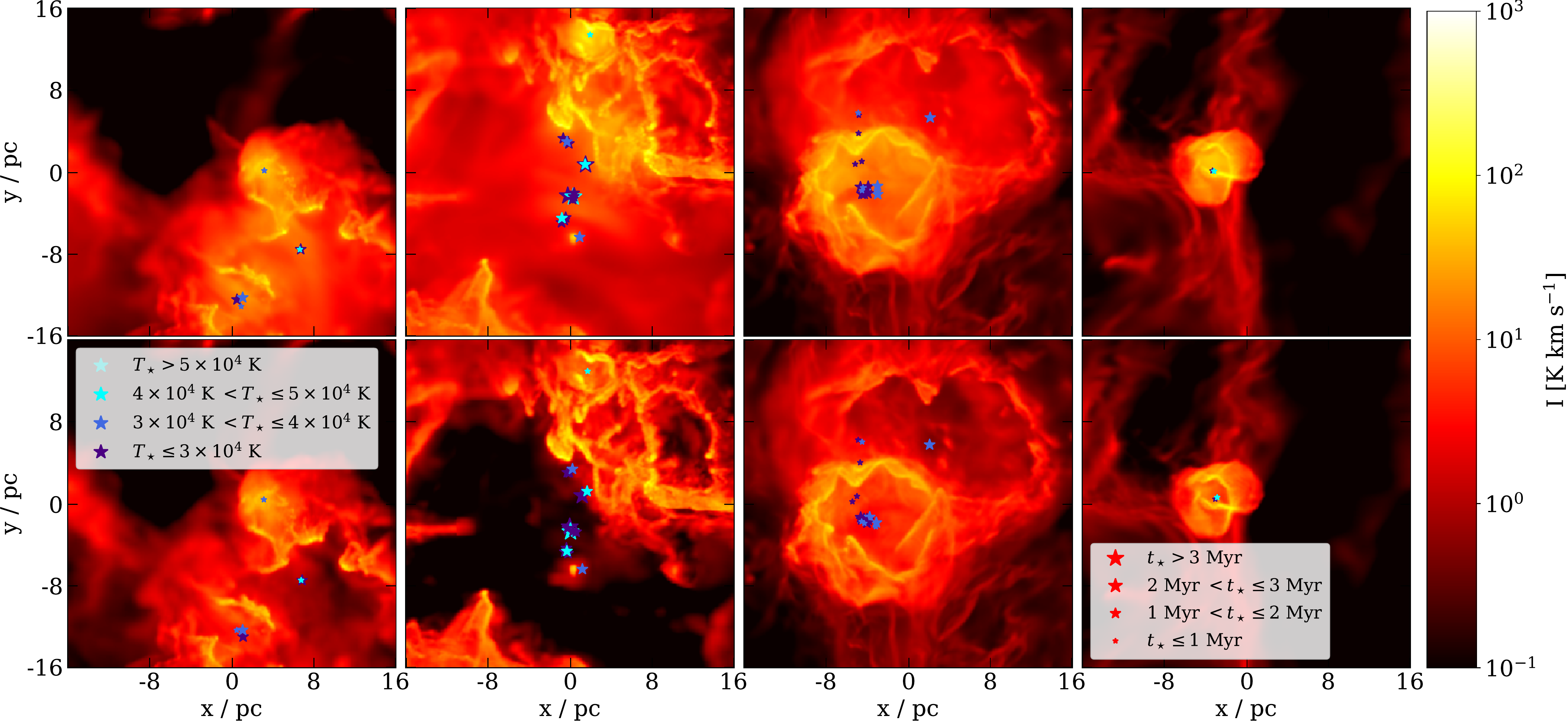}
    \caption{Examples of synthetic [CII] emission maps of expanding feedback bubbles, before (top row) and after the post-processing described in Section~\ref{sec:postprocess} to account for the conversion of C$^{+}$ into C$^{2+}$  (bottom row). The importance of the post-processing in reducing the [CII] intensity coming from the interior of the bubbles is evident, in particular for bubbles associated with older and hotter stars.}
    \label{fig:zoom_on_bubbles_comparison}
\end{figure*}

\begin{figure*}
    \centering
    \includegraphics[width=0.9\textwidth]{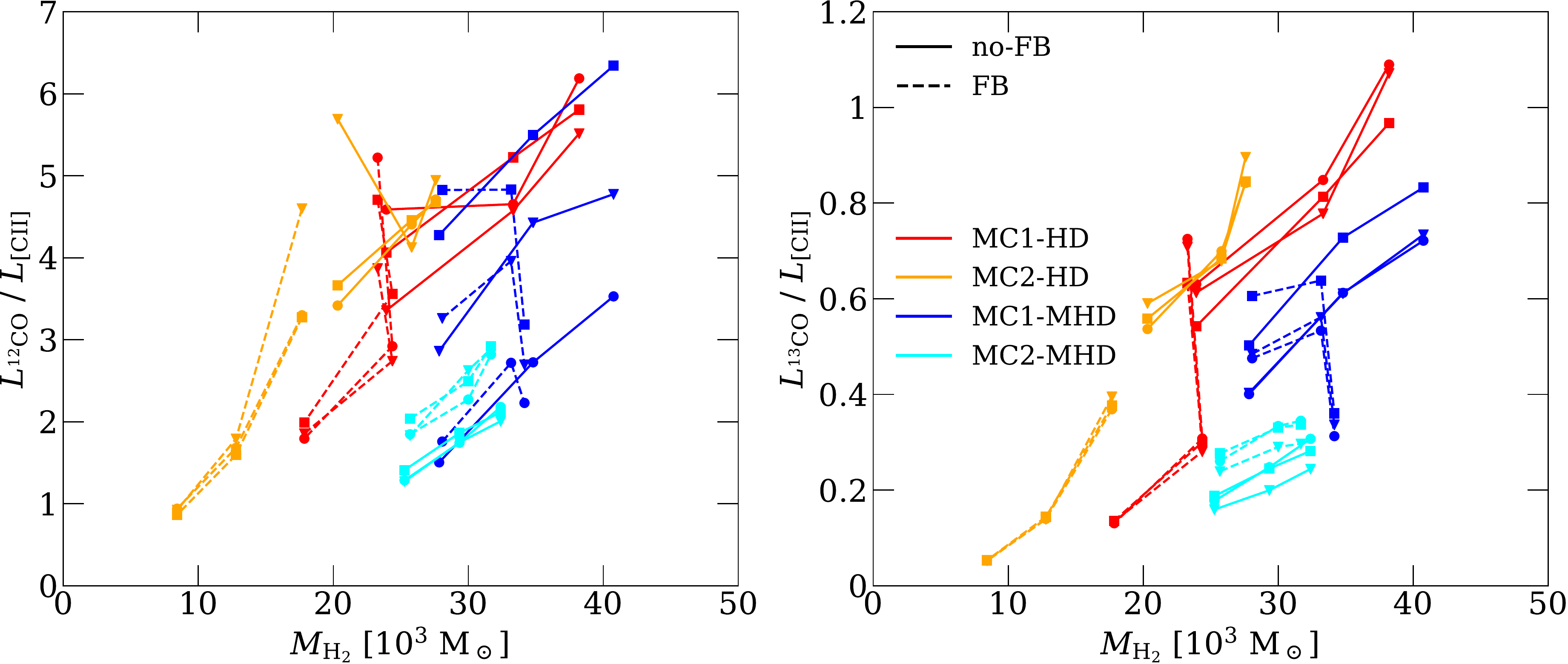}
    \caption{$L_\rmn{^{12}CO}/L_\rmn{[CII]}$ (left panel) and $L_\rmn{^{13}CO}/L_\rmn{[CII]}$ (right panel) as a function of the H$_2$ mass as opposed to Fig.~\ref{fig:line_ratio_global_fit} where it is plotted against $M_\rmn{H_2}/M_\rmn{H, tot}$. As the luminosity ratio is an intensive property of the clouds, while the H$_2$ mass is extensive, the relation shows an even larger scatter than the one against the H$_2$ mass fraction (see Fig.~\ref{fig:line_ratio_global_fit}).}
    \label{fig:line_ratio_vs_mass}
\end{figure*}

\begin{figure*}
    \centering
    \includegraphics[width=0.9\textwidth]{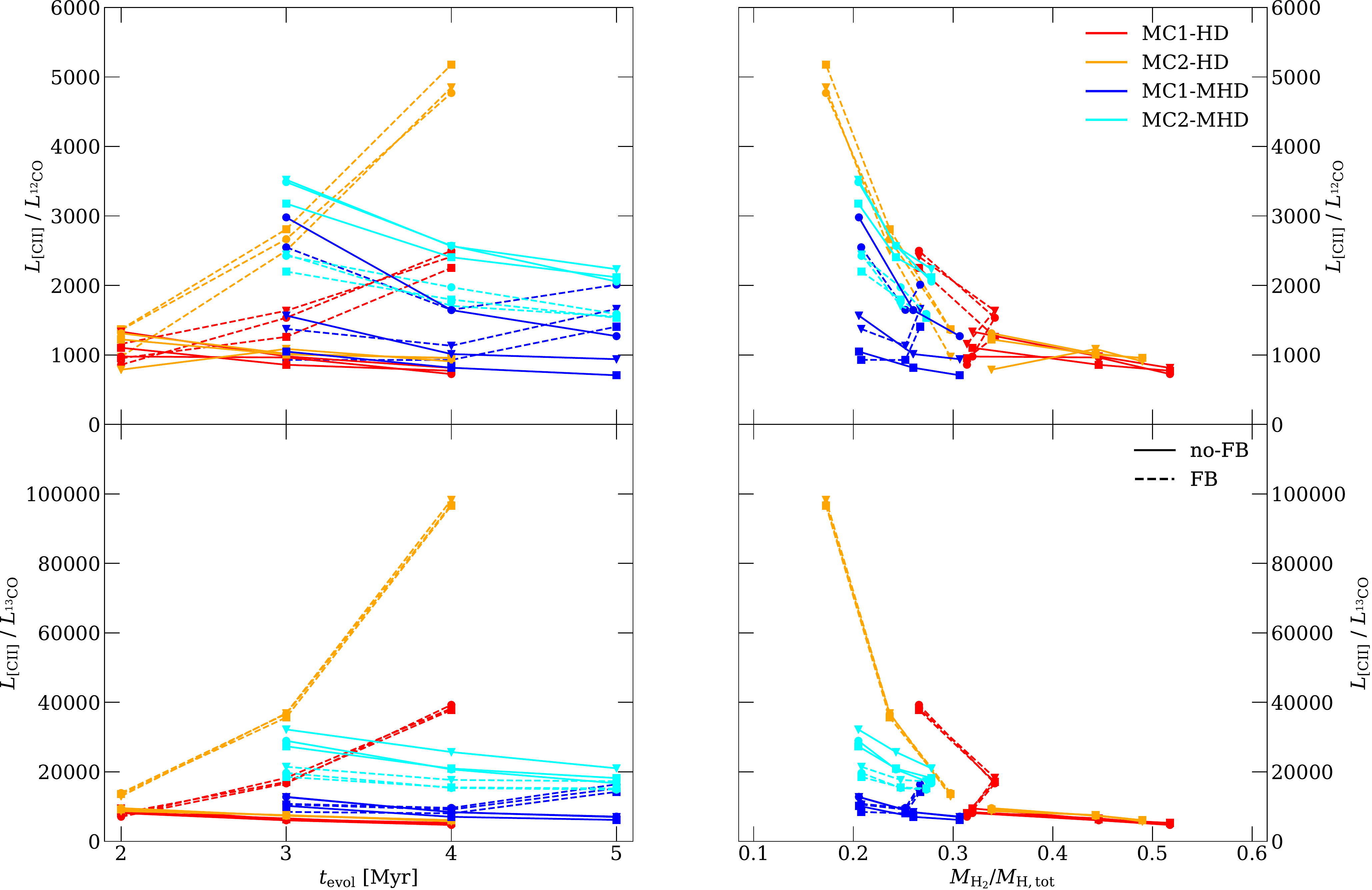}
    \caption{$L_\rmn{[CII]}/L_\rmn{^{12}CO}$ (top row) and $L_\rmn{[CII]}/L_\rmn{^{13}CO}$ (bottom row) as a function of $t_\rmn{evol}$ (left column) and $M_\rmn{H_2}/M_\rmn{H_{tot}}$ (right column). Luminosities are expressed in erg s$^{-1}$ as opposed to the usage of K~km~s$^{-1}$ in the main body of the paper.}
    \label{fig:line_ratio_erg}
\end{figure*}

\begin{figure*}
    \centering
    \includegraphics[width=0.9\textwidth]{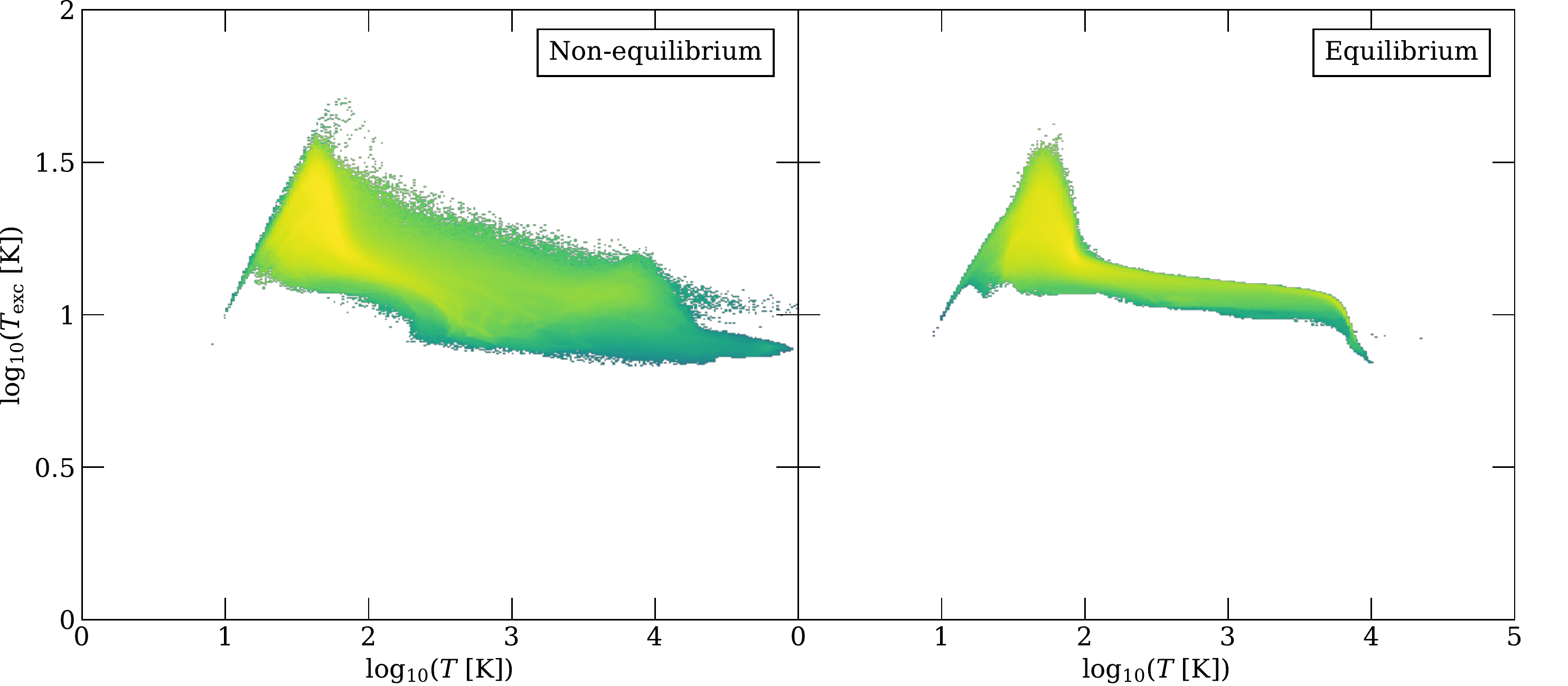}
    \caption{2D-PDF of excitation temperature $T_\rmn{ex}$ as a function of the gas temperature. The left-hand side plot represents MC1-HD-noFB at $t_\rmn{evol} = 2$ Myr with the chemistry evolved \textit{on-the-fly}, the right-hand side plot represents the same snapshot at steady state, i.e.~at $t_\rmn{chem} = 50$ Myr.}
    \label{fig:exc_temperature}
\end{figure*}

In the following, we show some additional plots which help in clarifying several aspects of the paper. In Fig.~\ref{fig:zoom_on_bubbles_comparison} we show the same expanding bubbles as in Fig.~\ref{fig:zoom_on_bubbles}, but now for the case with and without the post-processing for C$^{2+}$ (see Section \ref{sec:postprocess}) to allow for a direct comparison. The importance of the post-processing in removing the [CII] intensity coming from the interior of the bubbles is evident, in particular for bubbles associated with older and hotter stars.

In Fig.~\ref{fig:line_ratio_vs_mass} we show $L_\rmn{^{12}CO}/L_\rmn{[CII]}$ (left side) and $L_\rmn{^{13}CO}/L_\rmn{[CII]}$ (right side) as a function of the H$_2$ mass, instead of the H$_2$ mass fraction (see Fig.~\ref{fig:line_ratio_global_fit}). The correspondence of the line ratio with the H$_2$ mass is even weaker than with the H$_2$ mass fraction. This is due to the fact that the mass is an extensive quantity, whereas line ratios and mass fractions are intensive quantities.

In Fig.~\ref{fig:line_ratio_erg} we show the line ratios using units  of erg~s$^{-1}$ for the luminosity. This allows an easier comparison with some observational results e.g.~by \citet{Rollig2006} (see Section \ref{sec:variability_line_ratio}).

In Fig.~\ref{fig:exc_temperature} we show a 2D-PDF of the gas temperature and the excitation temperature for MC1-HD-noFB at $t_\rmn{evol} = 2$ Myr. The upper plot refers to the non-equilibrium (i.e., $t_\rmn{chem} = 0$), and the bottom plot refer to the equilibrium state ($t_\rmn{chem} = 50$ Myr) as discussed in Section~\ref{sec:equilibrium}. Due to the changes in the collisional partners (see Fig.~\ref{fig:mass_equilibrium}), the excitation temperature is lower at equilibrium, which explains why the [CII] luminosity decreases when evolving the chemistry to equilibrium. 

%%%%%%%%%%%%%%%%%%%%%%%%%%%%%%%%%%%%%%%%%%%%%%%%%%

% Don't change these lines
\bsp	% typesetting comment
\label{lastpage}
\end{document}